\documentclass[11pt, nofootinbib]{revtex4}
\usepackage{mathtools}
\usepackage{hyperref}
\usepackage{graphics}
\usepackage{subfigure}
\usepackage{epsfig}
\usepackage{amsmath,amsfonts}
\usepackage{amssymb}
\usepackage{mathrsfs}
\usepackage{color, xcolor}
\usepackage{soul}
\usepackage{ulem}
\usepackage{multirow, makecell}
\usepackage{threeparttable}
\usepackage{changes}

\begin{document}

\title{Gravitational wave polarization modes and stability analysis in Weyl geometry gravity}

\author{Yu-Zhi Fan$^{1,2}$}
\email{fanyzh2025@lzu.edu.cn}

\author{Xiao-Bin Lai$^{1,2}$}
\email{laixb2024@lzu.edu.cn}

\author{Yu-Qi Dong$^{1,2}$}
\email{dongyq2023@lzu.edu.cn}

\author{Yu-Xiao Liu$^{1,2}$}
\email{liuyx@lzu.edu.cn, corresponding author}

\affiliation{
	$^{1}$Lanzhou Center for Theoretical Physics,
	Key Laboratory of Theoretical Physics of Gansu Province, 
	Key Laboratory of Quantum Theory and Applications of MoE, 
	Gansu Provincial Research Center for Basic Disciplines of Quantum Physics,
	Lanzhou University, Lanzhou 730000, China \\
	$^{2}$Institute of Theoretical Physics $\&$ Research Center of Gravitation,
	School of Physical Science and Technology,
	Lanzhou University, Lanzhou 730000, China
}

\begin{abstract}
\textbf{Abstract:} We investigate the gravitational wave polarization modes and stability in Weyl geometry gravity within a Minkowski background. Our results indicate that the tensor sector consists of two standard modes propagating at the speed of light. Although the vector sector possesses a dynamical perturbation, it generates no polarization modes. The scalar sector, in contrast, features a mixture mode of breathing and longitudinal modes associated with a single scalar degree of freedom. The propagating solutions of this scalar mode exhibit superluminal propagation and can undergo intrinsic amplitude decay, both of which are driven by the background Weyl gauge field. We further discuss the observational detectability of this scalar mode. Our stability analysis reveals that, while the tensor and vector sectors are free from ghost and Laplacian instabilities, the scalar sector suffers from a ghost instability. These findings clarify the unique gravitational wave propagation characteristics in Weyl geometry gravity and provide theoretical foundations for testing the theory through future multi-messenger observations.
\end{abstract}

\maketitle

\section{Introduction}

\par Einstein's general relativity (GR) has achieved remarkable success in describing the gravitational interaction across a wide range of scales, from the solar system to the cosmological horizon. The recent direct detections of gravitational waves (GWs) by the LIGO and Virgo collaborations~\cite{LIGOScientific:2016aoc, LIGOScientific:2016emj, Abbott1, Abbott2} have opened a new window into the strong-field and dynamical regimes of gravity, providing unprecedented opportunities to test GR and its alternatives. Despite its remarkable achievements, GR faces conceptual and observational challenges, such as the singularity problem~\cite{cai2022spacetime}, the nature of dark matter~\cite{smith1936mass, Zwicky:1937zza} and dark energy~\cite{Peebles:2002gy}, and the long-standing quest for a consistent theory of quantum gravity~\cite{tHooft:1974toh, Goroff:1985th, Han:2005km}. These issues have motivated the exploration of modified gravity theories~\cite{clifton2012modified}.

\par Hermann Weyl proposed a novel geometric framework as a generalization of Riemannian geometry~\cite{Weyl:1918ib} in 1918, just two years after the publication of GR. In contrast to the metric-compatible and torsion-free spacetime of the Riemannian manifold, Weyl geometry relaxes the metric-compatibility condition by introducing a vector field $\omega_\mu$ (the Weyl gauge field) that characterizes the non-metricity of spacetime. Within this framework, the length of a vector is no longer invariant under parallel transport but becomes path-dependent, while angles between vectors remain preserved. Weyl's original purpose was to unify gravitation and electromagnetism, but this unified theory was criticized due to the prediction of a path-dependent clock rate, known as the second clock effect~\cite{Lobo:2018zrz}. Nevertheless, the geometric structure of this theory remains a subject of profound theoretical interest. Weyl geometry gravity has experienced a resurgence in the late twentieth century~\cite{Scholz:2017pfo}, and has been further developed in recent years~\cite{Ghilencea:2018thl, Ghilencea:2018dqd, Ghilencea:2019jux, Ghilencea:2019rqj, Ghilencea:2021jjl, Burikham:2023bil, Haghani:2023nrm, Oancea:2023ylb, Visa:2024fii, Khodadi:2025gtq, Lee:2024rjw, Sanomiya:2020svg, Duarte:2024zjb, Khodadi:2026zoi}, particularly in investigations concerning phenomenological and cosmological implications.

\par Gravitational waves provide a promising probe for testing gravity theories, particularly through their polarization modes, propagation speeds, and other characteristic properties. There are up to six polarization modes (the plus, cross, vector-$ x $, vector-$ y $, breathing, and longitudinal modes) in a general four-dimensional metric theory~\cite{Eardley:1973zuo}, while only the plus and cross modes exist in GR. Different gravity theories predict distinct sets of polarization modes as a consequence of the constraints imposed by their field equations. This feature enables tests of gravity theories by confronting theoretical predictions with observational constraints on GW polarizations. Over the past decade, the polarization content of GWs has been extensively investigated from both theory-specific~\cite{Liang:2017ahj, Hou:2017bqj, Gong:2018cgj, Wagle:2019mdq, Dong:2021jtd, Liang:2022hxd, Dong:2023xyb, Lai:2024fza, Fan:2024pex, Lai:2026vhe} and model-agnostic~\cite{Schumacher:2023jxq, Dong:2023bgt, Dong:2024zal, Dong:2025ddi, Dong:2025pyz} perspectives, providing a solid foundation for testing gravity theories using GW observations. A recent independent study~\cite{Ghilencea:2026jgp}, employing a de Sitter background and a different formulation of Weyl geometry gravity, has reported a polarization content distinct from ours. This difference may originate from the different choices of the background and the theoretical formulation adopted.

\par In recent years, the LIGO–Virgo–KAGRA Collaboration has performed a series of observational tests on GW polarizations, providing increasingly stringent constraints on possible additional polarization modes beyond GR~\cite{LIGOScientific:2017ous, LIGOScientific:2017ycc, LIGOScientific:2018czr, LIGOScientific:2018dkp, LIGOScientific:2019fpa, LIGOScientific:2021sio}. Beyond the current ground-based detector network, pulsar timing arrays~\cite{Sazhin:1978myk, Detweiler:1979wn} also provide an important and complementary approach for probing GW polarizations in the nanohertz band~\cite{Yunes:2013dva}. In this context, a preliminary indication of a scalar transverse polarization mode of GWs has recently been reported in the NANOGrav data set~\cite{Chen:2021wdo, Chen:2023uiz}. Looking ahead, space-borne detectors such as LISA~\cite{LISA:2017pwj}, Taiji~\cite{Hu:2017mde}, and TianQin~\cite{TianQin:2015yph} are expected to provide enhanced sensitivity to additional GW polarization modes in the millihertz band~\cite{Philippoz:2017ywb, Liu:2020mab, Wang:2021mou, Hu:2022byd, LISACosmologyWorkingGroup:2026zah}. In particular, LISA is estimated to be significantly more sensitive to longitudinal modes than to transverse modes at frequencies above roughly 0.06 Hz, while at lower frequencies it is equally sensitive to tensor and vector modes and slightly less sensitive to scalar modes~\cite{Tinto:2010hz, LISA:2024hlh, Zhang:2019oet, Zhang:2020khm}. Furthermore, future third-generation ground-based detector networks are expected to substantially improve the ability to distinguish additional GW polarization modes~\cite{Amalberti:2021kzh, ET:2025xjr}.

\par In addition to observational constraints, the theoretical viability of a gravity theory critically depends on its stability properties, since pathological instabilities may arise even in phenomenologically successful models. A primary concern is the ghost instability~\cite{Woodard:2006nt, DeFelice:2010gb}, which occurs when a propagating degree of freedom possesses a wrong-sign kinetic term, leading to a Hamiltonian unbounded from below and catastrophic vacuum decay. A well-known subclass is the Ostrogradsky ghost, which originates from nondegenerate higher-order time derivatives in the Lagrangian~\cite{Ostrogradsky:1850fid}. Another issue is the Laplacian instability~\cite{Bean:2007nx, DeFelice:2010gb}, associated with a negative squared sound speed of perturbations and the exponential growth of short-wavelength modes. These instabilities are closely tied to the dynamical structure of the Lagrangian, motivating a careful analysis of stability conditions required to avoid such pathologies~\cite{DeFelice:2016yws, Kase:2018aps, Clough:2022ygm, Zhang:2024cbw, DeFelice:2024dbj, vandeBruck:2025aaa, Lai:2025nyo}. For Weyl geometry gravity, the total action, consisting of the gravitational quadratic action and an effective matter action~\cite{Haghani:2023nrm, Visa:2024fii}, introduces additional dynamical degrees of freedom whose kinetic structure must be carefully examined via Hamiltonian analysis to determine the presence or absence of ghosts. Overall, the absence of ghost and Laplacian instabilities constitutes a fundamental consistency requirement for viable modified gravity theories.

\par In this paper, we investigate the GW polarization modes and the stability properties in Weyl geometry gravity. Working within a Minkowski background with a nonvanishing Weyl gauge field, we begin by performing a linear perturbation analysis and a scalar-vector-tensor decomposition to classify the perturbations into tensor, vector, and scalar sectors. We show that the tensor sector contains two standard modes propagating at the speed of light, while the vector sector generates no polarization modes despite possessing a dynamical perturbation. In contrast, the scalar sector features a mixture mode of breathing and longitudinal modes associated with a single scalar degree of freedom, which exhibits superluminal propagation and can undergo intrinsic amplitude damping in the propagating regime due to the background Weyl gauge field. Finally, we examine the theoretical consistency of the theory by carrying out a stability analysis and find that the tensor and vector sectors are free of ghost and Laplacian instabilities. Particularly, we reveal the presence of a ghost in the scalar sector by performing a Hamiltonian analysis. These results elucidate the distinctive GW signatures of Weyl geometry gravity and provide a theoretical basis for testing the theory with future GW observations.

\par The remainder of this paper is organized as follows. In Sec.~\ref{Sec.WGG}, we briefly review Weyl geometry gravity, including the geometric framework, the Weyl-invariant action, and the corresponding field equations. In Sec.~\ref{Sec.LPGI}, we derive the linear perturbation equations in a Minkowski background, and construct the gauge-invariant variables. In Sec.~\ref{Sec.GWP}, we analyze the polarization modes of GWs based on the gauge-invariant variables and discuss the properties and detectability of the scalar modes. In Sec.~\ref{Sec.Stab_TV}, we investigate the ghost and Laplacian stabilities in the tensor and vector sectors by deriving the second-order effective actions. In Sec.~\ref{Sec.Stab_S}, we examine the ghost instability in the scalar sector via a Hamiltonian analysis based on the second-order effective action. Finally, we summarize our main results and present concluding remarks in Sec.~\ref{Sec.Concl}.

\par We work in natural unit with $ c = \hbar = 1 $ and adopt the metric signature $ (-,+,+,+) $. Greek indices $ (\mu, \nu, \alpha, \beta, \cdots) $ range over spacetime components $ 0,1,2,3 $, while Latin indices $ (i, j, k, \cdots) $ are restricted to spatial components $ 1,2,3 $.

\section{An overview of Weyl geometry gravity}	\label{Sec.WGG}

\subsection{Weyl geometry}

\par Weyl geometry~\cite{Weyl:1918ib, Ghilencea:2018dqd} is a generalization of Riemannian geometry in which the spacetime metric $ g_{\mu\nu} $ is supplemented by an additional vector field $ \omega_\mu $, called the Weyl gauge field. The central idea is that the action is required to be invariant under (local) Weyl gauge transformations,
\begin{equation}	\label{WeylTrans}
	\begin{aligned}
		g_{\mu\nu} &\to g_{\mu\nu}' = \Omega^2(x) g_{\mu\nu},	\\
		\omega_\mu &\to \omega_\mu' = \omega_\mu - \frac1\alpha \partial_\mu \ln \Omega^2(x),	\\
		\phi &\to \phi' = \frac{\phi}{\Omega(x)},
	\end{aligned}
\end{equation}
where $ \Omega(x) > 0 $ is a conformal factor, $ \alpha $ is the Weyl gauge coupling, and $ \phi $ is an additional scalar field. 

\par The Weyl gauge field $ \omega_\mu $ measures the deviation from Riemannian geometry via a nonmetricity condition
\begin{equation} \label{nonmetricity}
	Q_{\lambda\mu\nu} \equiv \tilde \nabla_\lambda g_{\mu\nu} = - \alpha \omega_\lambda g_{\mu\nu},
\end{equation}
which indicates that angles between vectors are preserved under parallel transport but lengths of vectors may change depending on the path. Throughout this paper, quantities with a tilde are defined in Weyl geometry, and those without a tilde refer to their Riemannian counterparts. Note that both geometries share the same metric tensor, which is always denoted by $g_{\mu\nu}$; the difference between the two geometries lies in the connection rather than in the metric itself. The Weyl covariant derivative $ \tilde \nabla_\lambda $ is defined by the Weyl connection $ \tilde \Gamma^\lambda_{\ \mu\nu} $, whose relation to the Levi-Civita connection $ \Gamma^\lambda_{\ \mu\nu} $ can be found from Eq.~\eqref{nonmetricity}:
\begin{equation}
	\tilde \Gamma^\lambda_{\ \mu\nu} = \Gamma^\lambda_{\ \mu\nu} + \frac{\alpha}{2} (\delta^\lambda_{\ \mu} \omega_\nu + \delta^\lambda_{\ \nu} \omega_\mu - g_{\mu\nu} \omega^\lambda ).
\end{equation}

\par The curvatures in Weyl geometry are constructed by the Weyl connection $ \tilde \Gamma^\lambda_{\ \mu\nu} $:
\begin{equation}
	\begin{aligned}
		\tilde R_{\ \mu\nu\sigma}^\lambda &= \partial_\nu \tilde\Gamma_{\ \mu\sigma}^\lambda - \partial_\sigma \tilde\Gamma_{\ \mu\nu}^\lambda + \tilde\Gamma_{\ \nu\rho}^\lambda \tilde\Gamma_{\ \mu\sigma}^\rho - \tilde\Gamma_{\ \sigma\rho}^\lambda \tilde\Gamma_{\ \mu\nu}^\rho,	\\
		\tilde R_{\mu\sigma} &= \tilde R_{\ \mu\lambda\sigma}^\lambda,		\\
		\tilde R &= g^{\mu\nu} \tilde R_{\mu\nu}.
	\end{aligned}
\end{equation}
The relation between curvature scalars in Weyl and Riemannian geometry is given by
\begin{equation}	\label{tildeR}
	\tilde R = R - 3\alpha \nabla_\mu \omega^\mu - \frac32 \alpha^2 \omega^2,
\end{equation}
where $ \omega^2 = \omega_\mu \omega^\mu $. One can find that under Weyl gauge transformations~\eqref{WeylTrans}, $ \tilde R $ transforms covariantly, $ \tilde R' = \Omega^{-2}(x) \tilde R $, while $ \sqrt{-g} \tilde R^2 $ is invariant. 

\par Another important quantity invariant under Weyl gauge transformations~\eqref{WeylTrans} is the field strength $ \tilde F_{\mu\nu} $ of the Weyl gauge field, defined as
\begin{equation}	\label{tildeF}
	\tilde F_{\mu\nu} = \tilde\nabla_\mu \omega_\nu - \tilde\nabla_\nu \omega_\mu = \nabla_\mu \omega_\nu - \nabla_\nu \omega_\mu = \partial_\mu \omega_\nu - \partial_\nu \omega_\mu = F_{\mu\nu}.
\end{equation}
Here, the symmetry of the Weyl connection $ \tilde \Gamma^\lambda_{\ \mu\nu} = \tilde \Gamma^\lambda_{\ \nu\mu} $ has been used. It is easy to find that $ \tilde F_{\mu\nu} $ and $ \sqrt{-g} \tilde F^2 = \sqrt{-g} \tilde F_{\mu\nu}\tilde F^{\mu\nu} = \sqrt{-g} g^{\mu\alpha} g^{\nu\beta} \tilde F_{\mu\nu} \tilde F_{\alpha\beta} $ are both invariant under Weyl gauge transformations~\eqref{WeylTrans}.

\subsection{Action in Weyl geometry gravity} 

\par Motivated by Weyl’s original proposal in 1918~\cite{Weyl:1918ib}, and by its reformulations in recent years~\cite{Ghilencea:2018dqd, Ghilencea:2019jux, Ghilencea:2019rqj}, we start with the simplest Weyl quadratic action invariant under Weyl gauge transformations~\eqref{WeylTrans},
\begin{equation}	\label{ActionG}
	S_g = \int \left( \frac{1}{4!\xi^2} \tilde R^2 - \frac14 \tilde F^2  \right) \sqrt{-g} d^4 x,
\end{equation} 
where $ \xi $ is a dimensionless coupling constant. In addition to the gravitational action~\eqref{ActionG}, we also consider an effective matter action~\cite{Haghani:2023nrm, Visa:2024fii},
\begin{equation}	\label{ActionM}
	S_m = \int \mathcal L_m (\psi_m; g_{\mu\nu}, \omega_\mu, \phi)  \sqrt{-g} d^4 x.
\end{equation} 
Unlike the ordinary matter Lagrangian $ L_m $ that depends only on the matter fields $\psi_m$ and the metric $ g_{\mu\nu} $, the effective matter Lagrangian $ \mathcal L_m $ may, in principle, also depend on the extra fields $ \omega_\mu $ and $\phi$. The details of this generalization are presented in the next subsection.

\par The higher derivatives in $ \tilde R^2 $ can be recast into an additional scalar degree of freedom by introducing an auxiliary scalar field $ \phi $ such that $ \tilde R^2 = 2\phi^2 \tilde R - \phi^4  $. Then we get the gravitational action linear in $ \tilde R $:
\begin{equation}	
	S_g = \int \left[ \frac{1}{4!\xi^2} (2\phi^2 \tilde R - \phi^4) - \frac14 \tilde F^2  \right] \sqrt{-g} d^4 x,
\end{equation} 
which is equivalent to the initial action~\eqref{ActionG}, as follows from the equation of motion $ \phi^2 = \tilde R $ obtained by varying the action with respect to $ \phi $. This procedure is analogous to the standard scalar-tensor formulation commonly employed in $f(R)$ gravity~\cite{Sotiriou:2008rp, DeFelice:2010aj}, where higher-curvature terms are similarly linearized by introducing an auxiliary scalar field.

\par Using Eqs.~\eqref{tildeR} and~\eqref{tildeF} to replace $ \tilde R $ and $ \tilde F_{\mu\nu} $ by Riemannian geometrical quantities, and adding the matter action, we obtain the total action,
\begin{equation}	\label{ActionTot}
	S = \int \left[ \frac{\phi^2}{12\xi^2} \left( R - 3\alpha \nabla_\mu \omega^\mu - \frac32 \alpha^2 \omega_\mu\omega^\mu  \right) - \frac{\phi^4}{4! \xi^2} - \frac14 F^2 + \mathcal L_m \right] \sqrt{-g} d^4 x.
\end{equation}
This action will serve as the basis for our subsequent analysis. The gravitational sector explicitly exhibits the scalar–vector–tensor structure of Weyl geometry gravity in terms of Riemannian quantities, while the matter sector will be specified in the next subsection.

\subsection{Effective matter Lagrangian and self-consistency condition}	\label{IIC}

\par As will become clear below, the ordinary matter Lagrangian $ L_m (\psi_m; g_{\mu\nu}) $ is not generally consistent with the requirements associated with Weyl gauge invariance. We therefore consider a more general effective matter Lagrangian $ \mathcal L_m (\psi_m; g_{\mu\nu}, \omega_\mu, \phi) $ that is allowed to depend explicitly on the Weyl gauge field $ \omega_\mu $ and the scalar field $ \phi $~\cite{Berezin:2021iof, Berezin:2022odj, Berezin:2022phu}. In the following, we define
\begin{equation}	\label{TGF}
	\begin{aligned}
		\mathcal T_{\mu\nu}  &\equiv - \frac{2}{\sqrt{-g}}\frac{\delta\left( \sqrt{-g}\mathcal{L}_m \right) }{\delta g^{\mu\nu}},		\\
		\mathcal G_\mu & \equiv \frac{\delta\mathcal{L}_m }{\delta\omega^\mu}, 	\\
		\mathcal F  & \equiv \frac{\delta\mathcal{L}_m }{\delta\phi}. 	
	\end{aligned}
\end{equation}
\par It is remarkable that the gravitational sector of the action is already invariant under Weyl gauge transformations~\eqref{WeylTrans}, while the matter sector of the action is not necessarily gauge-invariant~\cite{Berezin:2021iof, Berezin:2022odj, Berezin:2022phu}. In fact, in order to get gauge-invariant equations of motion, it is sufficient to require only the variation of the matter action, $ \delta S_m $, to be gauge-invariant~\cite{Berezin:2021iof, Berezin:2022odj, Berezin:2022phu}. The variations of $ g^{\mu\nu} $, $ \omega_\mu $, and $ \phi $ induced by Weyl gauge transformations~\eqref{WeylTrans} are given by
\begin{equation}
	\begin{aligned}
		\delta g^{\mu\nu} &= - 2 \frac{\delta\Omega}{\Omega} g^{\mu\nu},		\\
		\delta \omega_\mu &= -\frac2\alpha \nabla_\mu \frac{\delta \Omega}{\Omega}, 	\\
		\delta \phi &= -\frac{\delta\Omega}{\Omega}\phi. 	\\
	\end{aligned}
\end{equation}
From the gauge-invariance of $ \delta S_m $, we obtain
\begin{equation}
	\begin{aligned}
		0 = \delta S_m
		&= -\frac12 \int \mathcal T_{\mu\nu} \delta g^{\mu\nu}\sqrt{-g} d^4x + \int \mathcal G^\mu \delta \omega_\mu\sqrt{-g} d^4x + \int \mathcal{F} \delta \phi \sqrt{-g} d^4x	\\
		&= \int \mathcal T_{\mu\nu} g^{\mu\nu} \frac{\delta\Omega}{\Omega} \sqrt{-g} d^4x - \frac2\alpha \int \mathcal G^\mu  \nabla_\mu \frac{\delta \Omega}{\Omega}\sqrt{-g} d^4x - \int \mathcal{F} \phi \frac{\delta\Omega}{\Omega} \sqrt{-g} d^4x	\\
		&= \int \bigg( \mathcal T + \frac2\alpha  \nabla_\mu \mathcal G^\mu - \mathcal{F} \phi  \bigg) \frac{\delta\Omega}{\Omega}\sqrt{-g} d^4x,	\\
	\end{aligned}
\end{equation}
where we have used the integration by parts, and $ \mathcal T \equiv g^{\mu\nu} \mathcal T_{\mu\nu} $. This gives the self-consistency condition:
\begin{equation}  \label{S-C cond0}
	\mathcal T + \frac2\alpha  \nabla_\mu \mathcal G^\mu - \mathcal{F} \phi = 0.
\end{equation}

\par In the following, for simplicity, we retain only the lowest-order terms constructed from the additional fields and adopt the following minimal realization of the effective matter Lagrangian~\cite{Haghani:2023nrm, Visa:2024fii}:
\begin{equation}	\label{EML}
	\mathcal L_m (\psi_m; g_{\mu\nu}, \omega_\mu, \phi) = L_m (\psi_m; g_{\mu\nu}) + \frac12 \gamma \omega^2 + \frac12 \sigma \phi^2,
\end{equation}
where $ \gamma $ and $ \sigma $ are coupling constants with dimensions of mass squared. Then we have the specific forms of Eq.~\eqref{TGF}:
\begin{equation}
	\begin{aligned}
		\mathcal T_{\mu\nu}  &= T_{\mu\nu} - \gamma ( \omega_\mu\omega_\nu -  \frac12 g_{\mu\nu} \omega^2  ) + \frac12 \sigma\phi^2 g_{\mu\nu},		\\
		\mathcal G_\mu &= \gamma\omega_\mu, 	\\
		\mathcal F &= \sigma\phi, 	\\
	\end{aligned}
\end{equation}
where $ T_{\mu\nu} $ is the energy-momentum tensor defined by the ordinary matter Lagrangian $ L_m $, with $ T \equiv g^{\mu\nu}  T_{\mu\nu} $ its trace. The self-consistency condition becomes
\begin{equation}	\label{S-C cond1}
	T + \gamma \left( \frac2\alpha \nabla_\mu\omega^\mu + \omega^2 \right) + \sigma \phi^2 = 0.
\end{equation}
From both Eqs.~\eqref{S-C cond0} and~\eqref{S-C cond1}, one can find that in the ordinary case $ \mathcal L_m = L_m $, the self-consistency condition reduces to the traceless condition, $ T = 0 $. Therefore, only matter fields with traceless energy-momentum tensors, such as the electromagnetic field, satisfy this condition, whereas most physical matter fields, e.g., a perfect fluid, do not. This restriction motivates the introduction of the generalized effective matter Lagrangian considered above.

\subsection{Field equations in Weyl geometry gravity} 

\par We consider the total action~\eqref{ActionTot} with the specific effective matter Lagrangian~\eqref{EML}. The variation of the action~\eqref{ActionTot} with respect to the metric $ g^{\mu\nu} $ gives the gravitational field equation,
\begin{equation}	\label{FEq_t}
	\begin{aligned}
		&\Phi G_{\mu\nu} + ( g_{\mu\nu}\square - \nabla_\mu\nabla_\nu ) \Phi + \frac{3\alpha}{2}  ( \omega_{\mu}\nabla_{\nu}\Phi + \omega_{\nu}\nabla_{\mu}\Phi - g_{\mu\nu}\omega^{\rho}\nabla_{\rho}\Phi ) 	\\
		&-\frac{3\alpha^{2}}{2} \Phi \left( \omega_{\mu}\omega_{\nu} - \frac12 g_{\mu\nu} \omega^2 \right)  + \frac14 \Phi^2 g_{\mu\nu} - 6 \xi^2 \left( F_{\mu\alpha}F_{\nu}^{\ \alpha} - \frac14 g_{\mu\nu} F^{2} \right) \\
		&= 6 \xi^2 \left[ T_{\mu\nu} -  \gamma \left( \omega_\mu\omega_\nu -  \frac12 g_{\mu\nu} \omega^2  \right) + \frac12 \sigma\Phi g_{\mu\nu} \right],
	\end{aligned}
\end{equation}
where we have introduced the Einstein tensor $ G_{\mu\nu} $, and denoted $ \Phi \equiv \phi^2 $.

\par Varying the action~\eqref{ActionTot} with respect to the Weyl gauge field $ \omega^\mu $ yields the vector field equation,
\begin{equation}	\label{FEq_v}
	\nabla^{\nu} F_{\mu\nu} + \frac{\alpha}{4\xi^2}(\alpha\omega_{\mu} \Phi - \nabla_{\mu} \Phi)  = \gamma \omega_\mu.
\end{equation}
Taking the covariant divergence of Eq.~\eqref{FEq_v}, one can obtain
\begin{equation}	\label{FEq_v_cd}
	\square \Phi = \alpha \nabla_\mu (\Phi \omega^\mu) - \frac{4\xi^2}{\alpha} \gamma \nabla_\mu \omega^\mu.
\end{equation}

\par Finally, varying the action~\eqref{ActionTot} with respect to the scalar field $ \phi $, we obtain the scalar field equation,
\begin{equation}	\label{FEq_s}
	R - 3\alpha\nabla_{\mu}\omega^{\mu}-\frac{3}{2}\alpha^{2}\omega^2 - \Phi = - 6 \xi^{2}\sigma.
\end{equation}

\par By using Eq.~\eqref{FEq_s} and the trace of Eq.~\eqref{FEq_t}, we obtain
\begin{equation}	\label{FEq_ss}
	\square \Phi = \alpha \nabla_\mu (\Phi \omega^\mu) + 2\xi^2 (T + \gamma \omega^2 + \sigma \Phi).
\end{equation}
This implies the coupling between the vector field $ \omega_\mu $ and the scalar field $ \Phi $. Here, one can find that the combination of Eqs.~\eqref{FEq_v_cd} and~\eqref{FEq_ss} yields exactly the self-consistency condition~\eqref{S-C cond1}. This result is not accidental; rather, it is expected from the Noether identity associated with the Weyl gauge symmetry of the action, as guaranteed by Noether's second theorem~\cite{Noether_1971, Ruzziconi:2019pzd}.

\par Now, we have set up the basic framework of Weyl geometry gravity needed for our subsequent analysis. We are therefore ready to investigate its GW polarization modes and the stability conditions.

\section{Linear perturbations and gauge invariants}	\label{Sec.LPGI}

\subsection{Background equations and linear perturbation equations}

\par We now adopt a homogeneous and isotropic Minkowski background for the perturbative analysis,
\begin{equation}	\label{BG}
	g_{\mu\nu} = \eta_{\mu\nu}, \quad \omega_\mu = \mathring \omega_\mu = (\omega_0, 0, 0, 0), \quad \Phi = \mathring\Phi,
\end{equation} 
where $ \eta_{\mu\nu} $ is the Minkowski metric, $ \mathring\Phi $ is a constant scalar field, and $ \mathring \omega_\mu $ is a constant vector field with only nonvanishing temporal component $ \omega_0 $ due to spatial isotropy of the background. Since the gravitational field equation~\eqref{FEq_t} contains a term $ \Phi G_{\mu\nu} $, a vanishing background scalar field would eliminate the Einstein tensor from the linear perturbation equation, preventing the recovery of GR. Therefore, consistency with the GR limit requires a nonvanishing background scalar field, $ \mathring\Phi \neq 0 $.

\par Regarding the polarization modes of GWs, it is reasonable to focus on the vacuum field equations with $ T_{\mu\nu} = 0 $. Substituting Eq.~\eqref{BG} into the field equations~\eqref{FEq_t}-\eqref{FEq_s} and combining the resulting equations, we arrive at the following background constraints:
\begin{equation}	\label{BGC}
	\alpha^2 \mathring\Phi = 4 \xi^2 \gamma, \qquad 
	\alpha^2 \omega_0^2 = 4 \xi^2 \sigma, \qquad 
	\gamma = 3\alpha^2 \sigma \neq 0.
\end{equation}
Now, the independent quantities among the six parameters $ \{ \alpha, \xi, \omega_0, \mathring \Phi, \gamma, \sigma \} $ can be chosen as $ \{ \alpha, \xi, \omega_0 \} $. For convenience, $\mathring \Phi$ will occasionally be used below, but it should be kept in mind that $ \mathring \Phi = 3 \alpha^2 \omega_0^2 $, as implied by Eq.~\eqref{BGC}. Besides, it is easy to examine that the self-consistency condition~\eqref{S-C cond1} is satisfied at background level, $ \gamma \omega_0^2 = \sigma \mathring\Phi $.

\par Next we consider perturbations around the Minkowski background,
\begin{equation}	\label{LPE}
	\begin{aligned}
		g_{\mu\nu} &= \eta_{\mu\nu} + h_{\mu\nu}, 	\\ 
		g^{\mu\nu} &= \eta^{\mu\nu} - h^{\mu\nu}, 	\\ 
		\omega_\mu &= \mathring\omega_\mu + \delta\omega_\mu, \\
		\omega^\mu &= \mathring\omega^\mu + \delta\omega^\mu - \mathring\omega_\nu h^{\mu\nu},	\\
		\Phi &= \mathring\Phi + \delta\Phi, 	\\
	\end{aligned}
\end{equation}
where the contravariant vector background and perturbation are defined as $ \mathring\omega^\mu \equiv \eta^{\mu\nu} \mathring\omega_\nu $ and $ \delta\omega^\mu \equiv \eta^{\mu\nu}\delta\omega_\nu $, respectively. Here and below, we use $ \eta^{\mu\nu} $ and $ \eta_{\mu\nu} $ to raise and lower the spacetime indices in linearized theory.

\par Substituting the expansions~\eqref{LPE} into field equations~\eqref{FEq_t}-\eqref{FEq_s} and keeping terms up to linear order, we obtain the linear perturbation equations in vacuum:
\begin{gather}
	\mathring\Phi G_{\mu\nu}^{(1)} + (\eta_{\mu\nu}\square - \partial_\mu\partial_\nu - \frac{3\alpha^2}{2} \mathring\omega_{\mu}\mathring\omega_{\nu})\delta\Phi + \frac{3\alpha}{2} \left( \mathring\omega_{\mu}\partial_{\nu}\delta\Phi + \mathring\omega_{\nu}\partial_{\mu}\delta\Phi - \eta_{\mu\nu}\mathring\omega^{\rho}\partial_{\rho}\delta\Phi \right)  = 0,	\label{LFEq_t} \\
	4 \xi^2 ( \square\delta\omega_\mu - \partial_\mu\partial_{\nu}\delta\omega^\nu  ) + \alpha\partial_{\mu}\delta\Phi - \alpha^{2} \mathring\omega_{\mu} \delta\Phi = 0,	\label{LFEq_v}	\\
	R^{(1)} - 3\alpha (\partial_{\mu}\delta\omega^{\mu}  - \mathring\omega_\nu \partial_\mu h^{\mu\nu} + \frac12 \mathring\omega^\mu \partial_\mu h ) - 3\alpha^{2} (\mathring\omega_{\mu}\delta\omega^{\mu} - \frac12 \mathring\omega_\mu\mathring\omega_\nu h^{\mu\nu}) - \delta\Phi   = 0,	\label{LFEq_s}
\end{gather}
where we have used the background conditions~\eqref{BGC} to obtain the simplified forms, and a superscript ``(1)" attached to a quantity denotes its linearized counterpart.

\subsection{Gauge invariants}

\par A common strategy in GW studies is to construct gauge-invariant variables~\cite{Bardeen:1980kt, Flanagan:2005yc, Caprini:2018mtu, Alves:2023rxs}. Based on their transformation properties under spatial rotations, the components of $  h_{\mu\nu} $ and $ \delta\omega_\mu $ can be decomposed into irreducible pieces~\cite{Bardeen:1980kt, Flanagan:2005yc}:
\begin{equation}	\label{3+1Decomp.}
	\begin{aligned}
		h_{tt} &= 2\psi_h,\\
		h_{ti} &= \beta_i + \partial_i \rho,	\\
		h_{ij} &= h_{ij}^{\text{TT}} + \frac13 H \delta_{ij} + \partial_{(i}\varepsilon_{j)} + (\partial_i\partial_j - \frac13 \delta_{ij}\nabla^2)\zeta,	\\
		\delta\omega_t &= \psi_\omega,	\\
		\delta\omega_i &= \mu_i + \partial_i \varphi,	\\
	\end{aligned}
\end{equation}
subject to the constraints
\begin{equation}	\label{3+1Decomp.T}
	\partial^i \beta_i = \partial^i \varepsilon_i  = \partial^i \mu_i = 0, \qquad \partial^i h_{ij}^{\text{TT}} = \delta^{ij} h_{ij}^{\text{TT}} = 0.
\end{equation}
This is known as the scalar-vector-tensor decomposition. In this scheme, the spatial vectors $h_{0i}$ and $\delta\omega_i$ are both decomposed into a transverse (divergence-free) part and a longitudinal (gradient) part, while the spatial tensor $h_{ij}$ is decomposed into a transverse-traceless part, a trace part, a longitudinal-transverse part, and a longitudinal-traceless part.

\par We consider the infinitesimal local coordinate transformation
\begin{equation}	\label{CoordTran}
	x'^\mu  = x^\mu + \xi^\mu(x) , \qquad \xi^\mu(x) = (A, B^i + \partial^i C),
\end{equation}
where we parameterize the displacement $ \xi^\mu $ using $ A, B_i $, and $ C $, with $ \partial_i B^i = 0 $. Under this coordinate transformation, the metric and vector perturbations transform as~\cite{Flanagan:2005yc, Maggiore:2007ulw, Bluhm:2007bd}
\begin{align}
	h'_{\mu\nu} &=  h_{\mu\nu} -  \partial_\mu\xi_\nu - \partial_\nu\xi_\mu  ,	 \\
	\delta \omega_\mu' &=  \delta \omega_\mu - \mathring\omega_\alpha \partial_\mu \xi^\alpha.
\end{align}
These lead to a set of detailed gauge transformations of the perturbative variables:
\begin{equation}	\label{GaugeTranABC}
	\begin{aligned}
		\psi_h \ &\to\  \psi_h-\dot{A}, \\
		\rho  \ &\to\  \rho-\dot{C}-A, \\
		H  \ &\to\   H - 2\nabla^2C, \\
		\zeta \ &\to\  \zeta-2C, \\
		\beta_{i} \ &\to \ \beta_i-\dot{B}_i, \\
		\varepsilon_{i}\ \ &\to \ \varepsilon_i-2B_i, \\
		h_{ij}^{\text{TT}}  \ &\to \ h_{ij}^{\text{TT}},	\\
		\psi_\omega  \ &\to \ \psi_\omega + \omega_0 \dot A ,	\\
		\mu_i \ &\to \  \mu_i,	\\
		\varphi  \ &\to \  \varphi + \omega_0 A ,	\\
		\delta\Phi \ &\to \ \delta\Phi.
	\end{aligned}
\end{equation}
Under these gauge transformations, one can construct a set of gauge invariants: 
\begin{equation}	\label{GaugeInvariants}
	\begin{aligned}
		h_{ij}^{\text{TT}} &,	\\
		\Xi_i &\equiv \beta_i - \frac12 \dot \varepsilon_i,	\\
		\Psi &\equiv -\psi_h + \dot \rho - \frac12\ddot \zeta,	\\
		\Theta &\equiv \frac13(H - \nabla^2\zeta),	\\
		\mu_i &,	\\
		\Sigma &\equiv \psi_\omega + \omega_0 (\dot\rho - \frac12  \ddot \zeta) ,	\\
		\Omega &\equiv \varphi + \omega_0 (\rho - \frac12  \dot \zeta),	\\
		\delta\Phi &,
	\end{aligned}
\end{equation}
together with the transverse constraint $ \partial^i \Xi_i = 0 $. Here, the notation ``$ \cdot $" above a quantity denotes $ \partial_0 $.

\par Now we have obtained eight gauge invariants: one tensor ($ h_{ij}^{\text{TT}} $), two vectors ($ \Xi_i $ and $ \mu_i $), and five scalars ($ \Psi, \Theta, \Sigma, \Omega $, and $ \delta\Phi $). There are totally eleven degrees of freedom in these gauge invariants. This indicates that we have eliminated the four gauge degrees of freedom by constructing the gauge invariants~\eqref{GaugeInvariants}.

\section{Polarization modes of GWs} \label{Sec.GWP}

\par In this section, we first introduce the definition of the polarization modes, and then analyze the polarization modes of GWs in Weyl geometry gravity with the gauge-invariant perturbation equations.

\subsection{Geodesic deviation equation and polarization modes}

\par As discussed in Sec.~\ref{IIC}, although the effective matter Lagrangian $ \mathcal L_m $ is introduced in a general form $ \mathcal L_m = \mathcal L_m (\psi_m; g_{\mu\nu}, \omega_\mu, \phi)$, we restrict ourselves to the minimal realization given by Eq.~\eqref{EML}, $\mathcal L_m = L_m + \frac12 \gamma \omega^2 + \frac12 \sigma \phi^2$. In this realization, the additional vector and scalar fields do not couple directly to the ordinary matter fields, and the ordinary matter Lagrangian $ L_m $ depends only on the matter fields and the metric, $ L_m = L_m ( \psi_m; g_{\mu\nu} ) $, i.e., the ordinary matter fields only directly couple to the metric. It should be noted that the motion of ordinary test particles is entirely determined by the ordinary matter Lagrangian $ L_m $, rather than $ \mathcal L_m $, and therefore obeys the same geodesic equation as in GR. If direct couplings of the ordinary matter fields to the Weyl gauge field and the scalar field are included, the geodesic equation may be modified, and new GW polarization modes may emerge~\cite{Dong:2025ddi}. Such couplings are beyond the scope of the present work.

\par
\begin{figure}[h]
	\centering
	\includegraphics[width=1\textwidth]{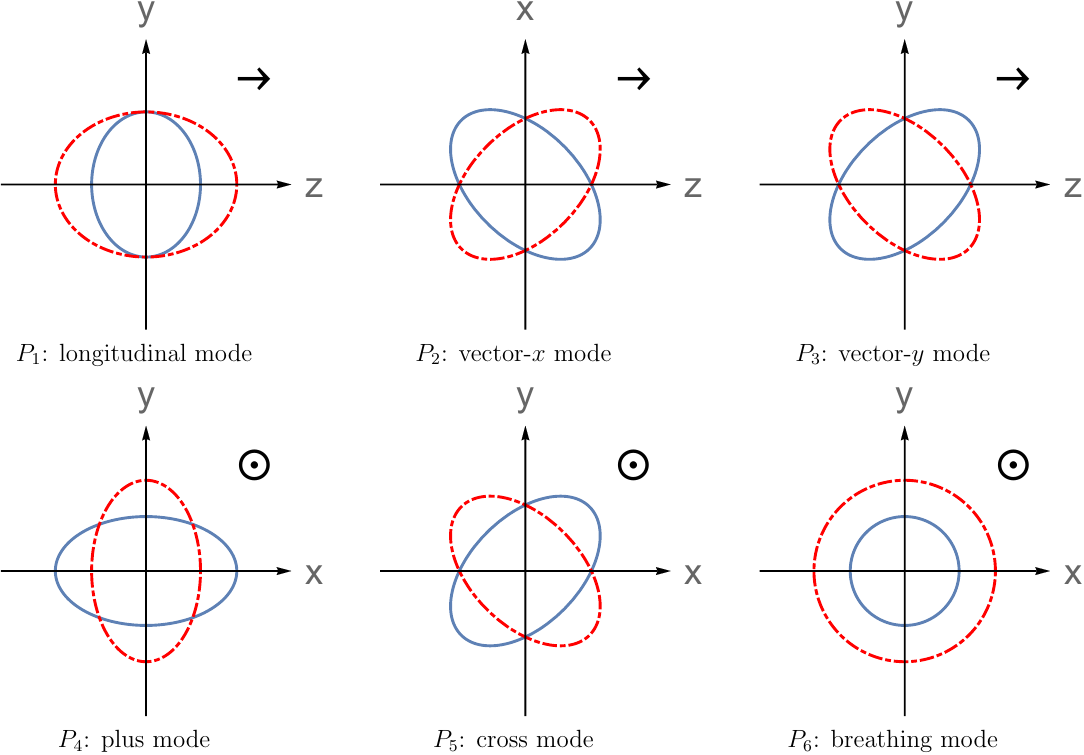}
	\caption{The relative motions of freely falling test particles induced by the six polarization modes of GWs~\cite{Eardley:1973zuo}. Initially, the test particles are positioned on a spherical shell. The GWs travel in the $ +z $ direction, as denoted by the icons in the upper-right corner of each subfigure. The solid and dashed curves both illustrate the relative positions of the test particles at their peak deviations, separated by a phase shift of half a cycle.}
	\label{Figure1}
\end{figure}

 The detectable effect of GWs with different polarization modes can be described by the geodesic deviation equation in the weak-field and low-speed limit:
\begin{equation} \label{GDEq_Ri0j0}
	\frac{d^2 \xi_i}{dt^2} = -R_{i0j0}^{(1)}\xi^j ,
\end{equation}
where $ \xi_i $ represents the spatial relative displacement between two adjacent test particles. The geodesic deviation equation~\eqref{GDEq_Ri0j0} shows that the relative motion of test particles is entirely determined by the $ (i0j0) $ components of the Riemann tensor, $ R_{i0j0} $. Therefore, the polarization modes of GWs can be described by $ R_{i0j0} $. We adopt the coordinate system in which the GWs propagate along the $ +z $ direction, and characterize the six independent polarization modes by $ R_{i0j0} $ as
\begin{equation} \label{R_i0j0_Polar.Mat.}
	R_{i0j0} = \frac12
	\begin{pmatrix}
		P_4 + P_6 &  P_5 & P_2 & \\
		P_5 & -P_4 + P_6 & P_3 &	\\
		P_2 &  P_3 & P_1 &	\\
	\end{pmatrix} .
\end{equation}
Here, $ P_1, \cdots, P_6 $ represent the longitudinal, vector-$ x $, vector-$ y $, plus, cross, and breathing modes, respectively. When GWs with a certain polarization mode propagate through freely falling test particles initially located on a spherical shell, their relative motions follow from the geodesic deviation equation~\eqref{GDEq_Ri0j0}, as illustrated in Fig.~\ref{Figure1}. 

\par We note that an alternative approach to characterizing the six polarization modes is the Newman-Penrose formalism~\cite{Newman:1961qr} widely used in the literature~\cite{Liang:2017ahj, Hou:2017bqj, Gong:2018ybk, Gong:2018cgj, Wagle:2019mdq, Dong:2021jtd, Hohmann:2018wxu, Gomes:2025ers}. For the null GWs, the six polarization modes are conveniently described by four Newman-Penrose quantities~\cite{Eardley:1973zuo}, whereas this description must be generalized for nonnull GWs~\cite{Hyun:2018pgn}. Since the scalar mode obtained below is nonnull and has a superluminal group velocity in the underdamped regime, we adopt the gauge-invariant approach throughout the present analysis.

\par The linear order of $ R_{i0j0} $ can be written in terms of gauge invariants~\cite{Flanagan:2005yc}:
\begin{equation}	\label{GI_R_i0j0}
	R_{i0j0} = -\frac12 \ddot h_{ij}^{\text{TT}} + \partial_{(i}\dot\Xi_{j)} + \partial_i\partial_j \Psi  - \frac12 \delta_{ij} \ddot\Theta.
\end{equation}
Using Eqs.~\eqref{R_i0j0_Polar.Mat.} and~\eqref{GI_R_i0j0}, we can find the relations between polarization modes and gauge invariants,
\begin{equation}		\label{Polar_GI}
	\begin{array}{lll}
		{ P_1 =  2\partial_3\partial_3\Psi -\ddot \Theta, \quad} & { P_2 = \partial_3\dot \Xi_1 , \quad	} & { P_3 = \partial_3\dot \Xi_2 , \quad }\\ 
		{ P_4 = -\ddot h_+,    } & { P_5 = -\ddot h_\times, \quad  } & { P_6 = -\ddot \Theta, } 	\\
	\end{array}
\end{equation}
with $ h_+ \equiv h_{11}^{\mathrm{TT}} = -h_{22}^{\mathrm{TT}} $ and $ h_\times \equiv h_{12}^{\mathrm{TT}} = h_{21}^{\mathrm{TT}} $. It is straightforward to verify that scalar, vector, and tensor perturbations generate the scalar modes $P_1$ and $P_6$, the vector modes $P_2$ and $P_3$, and the tensor modes $P_4$ and $P_5$, respectively. In the present analysis, all perturbative variables depend only on the coordinates $t$ and $z$. This restriction follows from the assumption that the GWs propagate along the $+z$ direction and that only wave solutions are considered. Consequently, spatial derivatives with respect to $x$ and $y$ vanish, and the $ z $ component of the transverse vector is identically zero, $ \Xi_3 = \mu_3 = 0 $. Besides, the modes $P_1$, $P_2$, and $P_3$ are not transverse, since the presence of derivatives along the propagation direction $z$ induces longitudinal variations.

\par Moreover, it should be emphasized that $R_{i0j0}$, which characterizes the polarization modes, depends exclusively on the gauge invariants constructed solely from the perturbations in metric, namely $h_{ij}^{\mathrm{TT}}$, $\Xi_i$, $\Psi$, and $\Theta$, as shown in Eq.~\eqref{GI_R_i0j0}. The remaining gauge invariants ($ \mu_i, \Sigma, \Omega $, and $ \delta\Phi $) do not enter $R_{i0j0}$ directly and therefore have no immediate effect on the polarization modes; however, they can influence them indirectly through the field equations, by sourcing or constraining $h_{ij}^{\mathrm{TT}}$, $\Xi_i$, $\Psi$, and $\Theta$.

\par Accordingly, we proceed by expressing the perturbation equations in terms of gauge invariants, so that the constraints imposed by the field equations on all gauge invariants can be analyzed, which in turn allows us to study the resulting polarization modes.

\subsection{Gauge-invariant perturbation equations and polarization modes}

\par Now we can use the gauge invariants~\eqref{GaugeInvariants} to rewrite the perturbation equations~\eqref{LFEq_t}-\eqref{LFEq_s} in a gauge-invariant form. This enables us to eliminate the gauge freedoms in the perturbation equations and reveal the physical ones, i.e., polarization modes. Besides, it is useful to decouple the $ ij $ components of Eq.~\eqref{LFEq_t} and the $ ti $ components of Eqs.~\eqref{LFEq_t} and~\eqref{LFEq_v} by taking the divergence of these equations.

\par First, we focus on the tensor perturbation equation~\eqref{LFEq_t}. The $ tt $ component of Eq.~\eqref{LFEq_t} written in terms of the gauge invariants is given by
\begin{equation}
	\nabla^2 (\mathring\Phi \Theta + \delta\Phi) - \frac{3\alpha}{2}  \omega_0 \delta\dot\Phi + \frac{3\alpha^2}{2} \omega_0^2  \delta\Phi  = 0.
\end{equation}
The gauge-invariant $ ti $ components of Eq.~\eqref{LFEq_t} can be decoupled into a scalar part and a transverse vector part:
\begin{align}
	\mathring\Phi \dot \Theta  + \delta\dot\Phi - \frac{3\alpha}{2} \omega_0 \delta\Phi   = 0,		\\
	\nabla^2 \Xi_i = 0.
\end{align}
The Poisson equation $ \nabla^2 \Xi_i = 0 $ leads to a trivial solution $ \Xi_i = 0 $ because we only consider the propagating GWs here. We also decouple the $ ij $ components of Eq.~\eqref{LFEq_t} in terms of the gauge invariants into the following independent parts:
\begin{align}
	\mathring\Phi (2\Psi + \Theta) + 2\delta\Phi = 0,	\\
	\square  h_{ij}^{\text{TT}} = 0.
\end{align}
The Klein-Gordon equation $ \square  h_{ij}^{\text{TT}} = 0 $ is the same as in GR, indicating the standard tensor polarization modes.

\par Then, consider the temporal component of the vector perturbation equation~\eqref{LFEq_v}. We obtain the gauge-invariant form:
\begin{equation}	\label{DampP0}
	4 \xi^2 \nabla^2 ( \Sigma - \dot \Omega  ) + \alpha\delta\dot\Phi - \alpha^{2}  \omega_0 \delta\Phi  = 0.
\end{equation}
The decoupled gauge-invariant spatial components of Eq.~\eqref{LFEq_v} gives
\begin{align}
	4 \xi^2 (\dot \Sigma - \ddot \Omega) + \alpha\delta\Phi   = 0,	\label{DampX0}	\\
	\square \mu_i  = 0.
\end{align}

\par Finally, we rewrite the scalar perturbation equation~\eqref{LFEq_s} in a gauge-invariant form as
\begin{equation}	\label{sclarEq0}
	\begin{aligned}
		 3 \ddot\Theta - 2\nabla^2 \Theta - 2 \nabla^2\Psi 
		 + 3\alpha \left[ \dot\Sigma - \nabla^2 \Omega + \omega_0  \left(\frac32 \dot\Theta - \dot\Psi \right)  \right]
		+ 3\alpha^2 \omega_0 ( \Sigma - \omega_0 \Psi ) - \delta\Phi  = 0.	     
	\end{aligned}
\end{equation}

\par We can now categorize these decoupled equations into three types---tensor, vector, and scalar equations---and analyze the corresponding polarization modes of GWs.

\par \textbf{(i) Tensor modes:}

\par The tensor equation
\begin{equation}	\label{WaveEqT}
	\square  h_{ij}^{\text{TT}} = 0
\end{equation}
indicates that there are two tensor modes propagating at the speed of light, as in GR.

\par \textbf{(ii) Vector modes:}

\par The vector equations
\begin{align}
	\Xi_i &= 0,	\\
	\square \mu_i  &= 0,	\label{WaveEqV}
\end{align}
show that one vector gauge invariant, $ \Xi_i $, vanishes, but the other, $ \mu_i $, is a propagating one. However, the gauge invariant $ \mu_i $ does not contribute to the polarization modes that are related to $ R_{i0j0} $. Although $ \mu_i $ represents propagating vector degrees of freedom, vector modes would exist only if the gauge invariant $ \Xi_i $ were nonvanishing, which is not the case here. Therefore, Weyl geometry gravity exhibits no vector polarization modes, despite containing two vector degrees of freedom.

\par \textbf{(iii) Scalar modes:}

\par The scalar equations are more complicated. We rearrange them and get a simpler set of equations:
\begin{gather}
	\nabla^2 (\mathring\Phi \Theta + \delta\Phi) - \frac{3\alpha}{2}  \omega_0 \delta\dot\Phi + \frac{3\alpha^2}{2} \omega_0^2  \delta\Phi  = 0, \label{Thetakk}	\\
	\mathring\Phi \dot \Theta  + \delta\dot\Phi - \frac{3\alpha}{2} \omega_0 \delta\Phi   = 0,	\label{Thetaw}	\\
	\mathring\Phi (2\Psi + \Theta) + 2\delta\Phi = 0,	\\
	\square \delta\Phi + \alpha \omega_0 \delta\dot \Phi = 0,	\label{DampP}\\
	4 \xi^2 ( \dot\Sigma - \ddot\Omega ) + \alpha \delta\Phi = 0,	\label{DampX}	\\
	3\square \delta\Phi + \frac{21}{2}\alpha^2\omega_0^2 \delta\Phi + 3\alpha \mathring\Phi( \dot\Sigma - \nabla^2\Omega ) + \frac{3}{2}\alpha^2 \mathring\Phi \omega_0 ( \omega_0 \Theta + 2\Sigma ) = 0.	\label{ScalarEq5}
\end{gather}

\par There are five variables $ \Theta, \Psi, \Sigma, \Omega, \delta\Phi $, but six equations. Actually, these equations are not independent, since Eq.~\eqref{DampP} can be obtained from Eqs.~\eqref{Thetakk} and~\eqref{Thetaw}. The number of the independent equations is the same as that of the variables. We are about to discuss the scalar modes encoded in these scalar equations in the next subsection.

\subsection{Discussion on scalar modes}	\label{Subsec. ScalarModes}

\par Since the scalar equations~\eqref{Thetakk}-\eqref{ScalarEq5} contain time derivatives of odd order, while spatial derivatives appear only at even order, we adopt a spatial plane-wave ansatz propagating in the $ +z $ direction,
\begin{equation}	\label{PlaneWave}
	\Pi (t,z) = \Pi (t, k) e^{ikz}, 
\end{equation}
where $ \Pi $ represents any one of the scalar gauge invariants $ \Theta, \Psi, \Sigma, \Omega $, and $ \delta\Phi $, and the time dependence is left arbitrary at this stage. 

{
\par With this ansatz~\eqref{PlaneWave}, the independent content of the scalar equations~\eqref{Thetakk}-\eqref{ScalarEq5} can be organized into two groups. The first group is given by
\begin{align}
	\delta\ddot\Phi - \alpha \omega_0 \delta\dot\Phi + k^2 \delta\Phi &= 0 ,	\label{Sol4}	\\
	\ddot\Omega + \alpha\omega_0 \dot\Omega + k^2 \Omega &= c_1 \delta\dot\Phi + c_2 \delta\Phi	,	\label{Sol5}
\end{align}
with the coefficients
\begin{equation}
	\begin{aligned}
		c_1 &\equiv -\frac{1}{4 \xi^2 k^2} ( \alpha^2 - \xi^2 ) \omega_0 + \frac{1}{3\alpha^2 \omega_0},	\\
		c_2 &\equiv -\frac{\alpha}{4 \xi^2 k^2} ( \xi^2 - \alpha^2 ) \omega_0^2 - \frac{1}{4\alpha\xi^2} ( 4\xi^2 - \alpha^2 ).
	\end{aligned}
\end{equation}
These two second-order differential equations~\eqref{Sol4} and~\eqref{Sol5} reveal the presence of two dynamical degrees of freedom in the scalar sector, represented by $ \delta\Phi $ and $ \Omega $, both of which require two initial conditions to determine a unique solution. This dynamical structure is consistent with the result of the stability analysis for the scalar sector presented later in Sec.~\ref{Sec.Stab_S}. 

\par The second group consists of relations that determine the remaining scalar variables:
\begin{align}
	\Theta &= \frac{1}{2\mathring\Phi k^2} (-3\alpha \omega_0 \delta\dot\Phi + (3\alpha^2\omega_0^2 - 2k^2) \delta\Phi ),	\label{Sol1}	\\
	\Psi &= \frac{1}{4\mathring\Phi k^2} (3\alpha\omega_0 \delta\dot\Phi - (3\alpha^2\omega_0^2 + 2k^2) \delta\Phi ),	\label{Sol2}	\\
	\Sigma &= \frac{\alpha}{4 \xi^2 k^2} ( \delta\dot\Phi - \alpha \omega_0 \delta\Phi ) + \dot\Omega .	
\end{align}
Once $\delta\Phi$ and $\Omega$ are determined, these relations fix $\Theta$, $\Psi$, and $\Sigma$ completely. In particular, $\Theta$ and $\Psi$ depend only on $\delta\Phi$, whereas $\Sigma$ depends on both $\delta\Phi$ and $\Omega$. Therefore, the second group introduces no additional independent scalar degrees of freedom. 
}

\par Now, we examine the scalar polarization modes in detail. According to Eq.~\eqref{Polar_GI}, the scalar modes depend directly only on $ \Theta $ and $ \Psi $, not on $ \delta\Phi $ or $ \Omega $. Nevertheless, $ \delta\Phi $ influences $ \Theta $ and $ \Psi $ through Eqs.~\eqref{Sol1} and~\eqref{Sol2}, and therefore affects the scalar modes indirectly, while $ \Omega $ has no impact on the polarization modes. From $ P_6 = - \ddot\Theta $ and $ P_1 = 2\partial_3\partial_3 \Psi - \ddot\Theta $, one can verify that both breathing and longitudinal modes do not vanish. Consequently, there is a mixture mode of breathing and longitudinal modes, corresponding to a single scalar degree of freedom.

\par We then turn our attention to the dynamical degree of freedom $ \delta\Phi $ that affects the polarization modes. We then adopt a full plane-wave ansatz,
\begin{equation}	\label{PWSol}	
	\delta\Phi(t,z) = \delta\Phi(w,k) e^{i(-wt+kz)} = \delta\Phi (w, k) e^{w_I t} e^{i (-w_R t + kz)},
\end{equation}
allowing for a complex frequency $ w = w_R + i w_I $ due to the presence of a first-order time derivative in the damped wave equation. Substituting this ansatz into Eq.~\eqref{Sol4}, {we obtain the dispersion relations for three distinct damping regimes:
\begin{alignat}{2}
	\text{Case (1):}\quad
	& w_R^\pm = \pm \frac{1}{2}
	\sqrt{4k^2-\alpha^2\omega_0^2},
	\qquad w_I = \frac{1}{2}\alpha\omega_0,
	&\qquad& \text{for}\quad
	4k^2>\alpha^2\omega_0^2,
	\label{case1}\\
	\text{Case (2):}\quad
	& w_R=0,
	\qquad w_I=\frac{1}{2}\alpha\omega_0,
	&\qquad& \text{for}\quad
	4k^2=\alpha^2\omega_0^2,
	\label{case2}\\
	\text{Case (3):}\quad
	& w_R=0,
	\qquad w_I^\pm=\frac{1}{2}\alpha\omega_0
	\pm\frac{1}{2}\sqrt{\alpha^2\omega_0^2-4k^2},
	&\qquad& \text{for}\quad
	4k^2<\alpha^2\omega_0^2.
	\label{case3}
\end{alignat}

\par Case (1) corresponds to the underdamped regime. To prevent exponential growth of the solution~\eqref{PWSol}, we require $w_I<0$, or equivalently, $\alpha\omega_0<0$. The solution~\eqref{PWSol} with dispersion relation~\eqref{case1} gives
\begin{equation}	\label{DampSol1}
	\delta\Phi(t,z)
	=
	A^{(1)} e^{w_I t} e^{-i w_R^+ t } e^{i kz} + B^{(1)} e^{w_I t} e^{-i w_R^- t } e^{i kz}.
\end{equation}
Here, $A^{(i)}$ and $B^{(i)}$ $(i=1,2,3)$ denote integration constants associated with the two independent solutions in Case $(i)$. The solution~\eqref{DampSol1} represents a plane wave whose amplitude decays with time. This decay is intrinsic to the field equations and does not arise from cosmological propagation effects, since the background spacetime is taken to be flat. The dispersion relation~\eqref{case1} implies that the propagation speed of the scalar modes is superluminal,
\begin{equation}
	v = \frac{dw_R}{dk} = \left[ 1 - \left( \frac{\alpha\omega_0}{2k} \right)^2 \right]^{-1/2} > 1.
\end{equation}}
This result shows that the deviation from the speed of light originates from the nonvanishing temporal component of the background vector field, $ \omega_0 $. As $ \omega_0^2 $ decreases or the GW wavenumber $ k $ increases, the propagation speed of the scalar modes decreases and approaches the speed of light. In other words, for a fixed background vector field, GWs with lower frequencies propagate at higher speeds. 

{
\par Cases (2) and (3) correspond to the critically damped and overdamped regimes, respectively. In Case (2), the two characteristic roots coincide, so the repeated-root solution takes the form
\begin{equation}
	\delta\Phi(t,z)
	=
	(A^{(2)} + B^{(2)} t)e^{w_I t}e^{ikz}.
\end{equation}
For Case (3), the two roots are distinct, and the general solution is
\begin{equation}
	\delta\Phi(t,z)
	=
	\left(A^{(3)} e^{w_I^+t} + B^{(3)} e^{w_I^-t}\right)e^{ikz}.
\end{equation}
In both cases, avoiding exponential growth requires the imaginary parts of the frequencies to be negative for any wavenumber satisfying $4k^2 \leq \alpha^2\omega_0^2$, which is equivalent to requiring $\alpha\omega_0<0$. Since $w_R=0$ in these two cases, the perturbations are non-oscillatory exponential modes rather than propagating oscillatory waves. Therefore, the discussion of the superluminal propagation speed based on a nonzero $w_R$ applies only to Case (1). Since our focus is on propagating oscillatory GWs, we restrict the subsequent analysis to the underdamped regime $4k^2 > \alpha^2\omega_0^2$ and impose $\alpha\omega_0<0$ to ensure decay.

\par In addition to $\delta\Phi$, it is necessary to clarify the behavior of the other scalar degree of freedom, $\Omega$. The homogeneous parts of Eqs.~\eqref{Sol4} and~\eqref{Sol5} differ only in the signs of their damping terms. The homogeneous solution for $\Omega$ therefore grows under the condition $\alpha\omega_0<0$, which ensures the decay of $\delta\Phi$, the degree of freedom contributing to the scalar polarization mode. This does not imply that every solution for $\Omega$ grows, since a non-growing solution can still be constructed by choosing appropriate initial conditions for $\Omega$. Using Eq.~\eqref{Sol4}, one can verify that Eq.~\eqref{Sol5} admits the particular solution
\begin{equation}
	\Omega_{\mathrm p}
	=\frac{-c_2\delta\dot\Phi+(k^2c_1+\alpha\omega_0c_2)\delta\Phi}{2\alpha\omega_0 k^2},
	\qquad 
	\dot\Omega_{\mathrm p}
	=\frac{c_1\delta\dot\Phi+c_2\delta\Phi}{2\alpha\omega_0}.
\end{equation}
For $\alpha\omega_0<0$, the decay of $\delta\Phi$ and $\delta\dot\Phi$ likewise leads to the decay of $\Omega_{\mathrm p}$ and $\dot\Omega_{\mathrm p}$. The general solution can be written as $\Omega = \Omega_{\mathrm p} + \Omega_{\mathrm h}$, where the homogeneous part $\Omega_{\mathrm h}$ satisfies $\ddot\Omega_{\mathrm h} + \alpha\omega_0\dot\Omega_{\mathrm h} + k^2\Omega_{\mathrm h} = 0$. By choosing the initial conditions of $\Omega$ to coincide with those of the particular solution,
\begin{equation}	\label{IniCon}
	\Omega(t_0)=\Omega_{\mathrm p}(t_0),\qquad
	\dot\Omega(t_0)=\dot\Omega_{\mathrm p}(t_0),
\end{equation}
we obtain $\Omega_{\mathrm h}(t_0)=\dot\Omega_{\mathrm h}(t_0)=0$. By uniqueness of the solution to the homogeneous initial-value problem, $\Omega_{\mathrm h}(t)$ then vanishes identically, which is equivalent to setting both homogeneous integration constants to zero. The evolution therefore remains on the decaying solution $\Omega=\Omega_{\mathrm p}$. This decaying solution corresponds to a special choice of the initial conditions for $\Omega$. For generic initial conditions, the homogeneous part $\Omega_{\mathrm h}$ is nonvanishing and grows under $\alpha\omega_0<0$. Thus, this condition guarantees the decay of $\delta\Phi$, but the absence of a growing $\Omega$ mode additionally requires the initial-condition choice~\eqref{IniCon}.
}

\subsection{Detectability of scalar modes}

\par Now, we present an illustrative estimate of the detectability of scalar GWs within the underdamped regime~\eqref{case1} under the condition $ \alpha\omega_0 < 0 $. Although $\delta\Phi$ does not enter the geodesic deviation equation directly, Eqs.~\eqref{Sol1} and~\eqref{Sol2} show that the gauge-invariant metric perturbations $\Theta$ and $\Psi$, which determine the scalar polarization modes, inherit the same damping factor $e^{w_I t}$ from $\delta\Phi$.

\par Consider a GW source at a typical distance, for example the luminosity distance of GW150914~\cite{LIGOScientific:2016aoc,LIGOScientific:2016emj}, $ d_L \sim 410 $ Mpc $ \sim 1.3 \times 10^9 $ light years. Assuming that the deviation of the propagation speed from the speed of light is small, the travel time to the detector can be estimated as $ t_{\mathrm{typ}} \sim 1.3 \times 10^9 $ years. The scalar modes are possible to be detected before the amplitude decays to $ 1/e $ of its initial value, only if the parameters satisfy
\begin{equation}	\label{ParaCon}
	-m_s \lesssim w_I = \frac12 \alpha \omega_0 < 0, \qquad  m_s \equiv \frac{1}{t_{\mathrm{typ}}}  \sim 1.6 \times 10^{-32} \ \mathrm{eV},
\end{equation}
where the boundary $ m_s $ characterizes the damping rate of the scalar modes and sets an effective energy scale associated with the decay of its amplitude. Otherwise, the amplitude may be too small to be detected upon arrival at the detector.

\par If we further adopt the boundary $ m_s $ as a representative estimate for the parameters ($ \alpha \omega_0 \sim -2 m_s \sim -3.2\times 10^{-32} \ \mathrm{eV} $), the difference in the arrival times between the tensor and the scalar modes can be estimated accordingly,
\begin{equation}
	\Delta t = d_L - \frac{d_L}{v} \approx \frac{\alpha^2\omega_0^2}{32 \pi^2 f^2} d_L \sim 4.9 \times 10^{-22} \ \mathrm{s},
\end{equation}
where we have used $ k \approx 2\pi f $ and taken $ f = 25 $ Hz as a typical frequency in the sensitive band of ground-based detectors (with these parameter estimates satisfying $4k^2 \gg \alpha^2\omega_0^2$, corresponding to the underdamped regime of Case (1)). This extremely small difference in the arrival times implies that, if one aims to detect the scalar GWs before their amplitudes decay to $ 1/e $ (i.e., under the parameter condition~\eqref{ParaCon}), the scalar and tensor GWs have to arrive almost simultaneously, with their signals overlapping. Otherwise, if we do not restrict to the condition~\eqref{ParaCon}, the scalar GWs may arrive noticeably earlier than the tensor GWs, but their amplitudes would be too small to be detected.

\par To conclude, the possibly detectable scalar-mode signal would arrive almost simultaneously with the tensor-mode signal, while scalar modes that arrive significantly earlier are likely too weak to be observed. Therefore, within the context of testing Weyl geometry gravity, a very long searching window in observation time is not required to examine the extra polarizations. However, in practice, since one generally aims to test multiple gravity theories, a longer observation time window would still be necessary~\cite{Schumacher:2023jxq, Schumacher:2025plq}.

\section{Stability analysis for tensor and vector perturbations} \label{Sec.Stab_TV}

\par In this section, we examine the ghost and Laplacian stabilities of tensor and vector perturbations through their dynamical second-order actions. We start from the action~\eqref{ActionTot} with the effective matter Lagrangian~\eqref{EML}, and consider the vacuum case without the source, $ L_m = 0 $.

\subsection{Tensor perturbations}

\par Consider the line element with only tensor perturbations,
\begin{equation}
	ds^2 = - dt^2 + ( \delta_{ij} +  h_{ij}^{\mathrm{TT}} ) dx^i dx^j.
\end{equation}
Using the background constraints~\eqref{BGC} and integration by parts, we obtain the second-order action 
\begin{equation}	\label{SoAT}
	S^{(2)}_\mathrm{T} =  \int  \frac{\mathring \Phi}{24\xi^2} \left( \dot h_+^2 + \dot h_\times^2 - \partial_i h_+\partial^i h_+ - \partial_i h_\times\partial^i h_\times \right)    d^4 x.
\end{equation}
The absence of ghost instability requires 
\begin{equation}
	\mathring \Phi > 0,
\end{equation}
which is natural due to the definition $ \Phi = \phi^2 $. The second-order action~\eqref{SoAT} also reveals that the GWs propagate at the speed of light, and there is no Laplacian instability. The variation of the action~\eqref{SoAT} with respect to $ h_+ $ and $ h_\times $ gives exactly the wave equation of the tensor modes~\eqref{WaveEqT}.

\subsection{Vector perturbations}

\par Now, we consider only the vector perturbations in~\eqref{3+1Decomp.}. The perturbed line element and vector field are
\begin{align}
	ds^2 &= - dt^2 + 2\beta_i dt dx^i + ( \delta_{ij} +  \partial_{(i}\varepsilon_{j)} ) dx^i dx^j,	\\
	\omega_\mu &= (\omega_0, \mu_i).
\end{align}
Under the background constraints~\eqref{BGC}, the second-order action induced by vector perturbations can be written in a gauge-invariant form,
\begin{equation}	\label{SoAV}
	S^{(2)}_\mathrm{V} = \int \left[ \frac{\mathring\Phi}{24 \xi^2} \partial_j\Xi_i\partial^j\Xi^i  + \frac{1}{2} \left( \dot\mu_i \dot\mu^i - \partial_j\mu_i\partial^j\mu^i \right) \right] d^4 x,
\end{equation}
where we have discarded the total derivatives after integrating by parts.

\par It is straightforward to identify $ \Xi_i $ as a nondynamical variable, since no time derivative of $ \Xi_i $ appears in the action. Varying the action~\eqref{SoAV} with respect to $ \Xi_i $ therefore yields the constraint equation $ \nabla^2 \Xi_i = 0 $. As we are only concerned with the dynamical degrees of freedom, we may eliminate this variable using its constraint equation, and obtain a second-order effective action containing only the dynamical variable,
\begin{equation}	\label{SoAV2}
	S^{(2)}_\mathrm{V} = \int \frac12 \left( \dot\mu_i \dot\mu^i - \partial_j\mu_i\partial^j\mu^i \right) d^4 x.
\end{equation}
This is a healthy action free from both ghost and Laplacian instabilities, since the kinetic and gradient terms in the action~\eqref{SoAV2} carry constant coefficients with the correct signs. The vector sector contains two dynamical degrees of freedom encoded in $ \mu_i $, whose equation of motion obtained from the action coincides with the previously derived wave equation~\eqref{WaveEqV}.

\section{Stability analysis for scalar perturbations}	\label{Sec.Stab_S}

\par In this section, we derive the second-order effective action for scalar perturbations after gauge fixing, and then we perform a Hamiltonian analysis to demonstrate the presence of a ghost in the scalar sector.

\subsection{Second-order action and gauge issues}

\par Let us focus on the scalar perturbations in~\eqref{3+1Decomp.}, which form the most involved sector. These perturbations appear in all relevant fields: the metric, the vector field, and the scalar field. The perturbed line element and vector field take the form
\begin{equation}
	\begin{aligned}
		ds^2 &= (-1+2\psi_h) dt^2 + 2 \partial_i \rho dt dx^i + \left( (1 + \frac{1}{3} H) \delta_{ij}  + (\partial_i\partial_j - \frac13 \delta_{ij} \nabla^2 ) \zeta \right) dx^i dx^j,	\\
		\omega_\mu &= ( \omega_0 + \psi_\omega, \partial_i \varphi),	\\
		\Phi &= \mathring\Phi + \delta\Phi.
	\end{aligned}
\end{equation}
In total, there are seven scalar perturbations, $ \psi_h, \rho, H, \zeta, \psi_\omega, \varphi $, and $ \delta\Phi $. Substituting these perturbations into the full action~\eqref{ActionTot} and using integration by parts along with the background equations~\eqref{BGC}, we obtain the second-order action
\begin{equation}	\label{SoAS}
	\begin{aligned}
		S^{(2)}_\mathrm{S} 
		&= \int \frac{1}{72 \xi^2} \bigg[ - \mathring\Phi \dot H^2 - 6\dot H \delta\dot\Phi - 9\alpha( \omega_0H + 2 \omega_0 \psi_h + 2\psi_\omega ) \delta\dot\Phi 	\\
		&\quad\, + \frac13 \mathring\Phi \partial_i H \partial^i H + 36 \xi^2 \partial_i \psi_\omega \partial^i \psi_\omega + 4 \partial_i H \partial^i \delta\Phi - 4 \mathring\Phi \partial_i H\partial^i \psi_h - 12 \partial_i \delta\Phi\partial^i \psi_h 	\\
		&\quad\,  - \frac{9\alpha^2\omega_0^2}{\mathring\Phi} \delta\Phi^2 + 18\alpha^2\omega_0 (\omega_0 \psi_h + \psi_\omega) \delta\Phi + \mathcal L_\varphi + \mathcal L_\rho + \mathcal L_\zeta  \bigg] d^4 x.
	\end{aligned}
\end{equation}
Here, $ \mathcal L_\zeta $ collects all terms involving $ \zeta $. The remaining terms containing $ \varphi $ and $ \rho $ are grouped into $ \mathcal L_\varphi $ and $ \mathcal L_\rho $, respectively. The explicit expressions for $ \mathcal L_\varphi, \mathcal L_\rho $, and $ \mathcal L_\zeta $ are given by
\begin{equation}
	\begin{aligned}
		\mathcal L_\varphi &= 18 ( 2 \xi^2 \partial_i\dot\varphi \partial^i\dot\varphi - 4 \xi^2 \partial_i \psi_\omega \partial^i \dot\varphi + \alpha \partial_i \delta\Phi \partial^i \varphi ),		\\
		\mathcal L_\rho &= -2 \partial_i ( 2 \mathring\Phi \dot H + 6 \delta\dot\Phi - 9 \alpha \omega_0 \delta\Phi ) \partial^i \rho, 		\\
		\mathcal L_\zeta &= \mathring\Phi \bigg( \nabla^2 \dot\zeta \nabla^2 \dot\zeta - 4 \nabla^2\rho \nabla^2\dot\zeta + \frac13 \partial_i\nabla^2 \zeta \partial^i\nabla^2 \zeta + 4\nabla^2 \Big( \frac{\delta\Phi}{\mathring\Phi} - \psi_h + \frac16 H \Big) \nabla^2 \zeta  \bigg).
	\end{aligned}
\end{equation}

\par Moreover, one can verify that there are gauge degrees of freedom in the scalar perturbations originating from the invariance of the linearized theory under infinitesimal local coordinate transformations~\eqref{CoordTran} with $ \xi^\mu = (A, \partial^i C) $. Therefore, by selecting appropriate values for $A$ and $ C $, we can eliminate specific scalar variables via the gauge transformations in Eq.~\eqref{GaugeTranABC}. In particular, the function $ C $ can be used to eliminate $ \zeta $, while $ A $ may be employed to fix either $ \rho $ or $ \varphi $. This leads to two possible gauge choices,
\begin{align}
	\text{Gauge 1:}\quad &\zeta = \rho = 0,		\label{Gauge1}	\\
	\text{Gauge 2:}\quad &\zeta = \varphi = 0.	\label{Gauge2}
\end{align}

\par In the following, we focus only on the first gauge choice~\eqref{Gauge1}. The analysis in the second gauge~\eqref{Gauge2} follows the same steps and leads to the same physical conclusions, which will thus not be presented separately.


\par Under the first gauge condition $ \zeta = \rho = 0 $, the second-order action~\eqref{SoAS} reduces to a form with $ \mathcal L_\zeta = \mathcal L_\rho = 0 $. In this action, it can be noticed that the variables $ \psi_\omega $ and $ \psi_h $ are nondynamical, as no time derivatives of these fields appear. Since our interest is restricted to the dynamical effects, the nondynamical variables should be eliminated using their constraint equations in order to obtain an effective action containing only the dynamical variables.

\par Varying the second-order action~\eqref{SoAS} with respect to $ \psi_\omega $ and $ \psi_h $, respectively, yields their constraint equations in Fourier space,
\begin{align}
	\mathcal E_{\psi_\omega} &\equiv \alpha \delta\dot\Phi - \alpha^2 \omega_0 \delta\Phi + 4 \xi^2 k^2 ( \dot\varphi - \psi_\omega ) = 0,	\label{CEqpso}	\\
	\mathcal E_{\psi_h} &\equiv 3 \alpha \omega_0 \delta\dot\Phi + (2 k^2 - 3 \alpha^2 \omega_0^2) \delta\Phi  + 2\alpha^2 \omega_0^2 k^2 H = 0,	\label{CEqpsh}
\end{align}
where $ k^2 = \delta^{ij} k_i k_j $ with $ k_i $ a 3-dimensional wave vector. 

\par Using the first constraint equation~\eqref{CEqpso}, one can solve for $ \psi_\omega $ and substitute the solution back into the action, thereby eliminating this variable. The resulting second-order action in Fourier space is
\begin{equation}	\label{SoAS1}
	\begin{aligned}
		S^{(2)}_\mathrm{S} 
		&=  \int  \mathcal{L}_\mathrm{S}   d^4 x = - \frac{1}{24 \xi^2} \int  \Big[  \mathcal{L}_0 (\delta\dot\Phi, \delta\Phi; \dot H, H; \dot\varphi, \varphi) + 2 \mathcal E_{\psi_h} \psi_h  \Big] d^4 x,	\\
		\mathcal{L}_0  
		&=  \frac{3\alpha^2}{4\xi^2 k^2} \delta\dot\Phi^2 + \alpha^2 \omega_0^2 \dot H^2 + 2 \delta\dot\Phi \dot H + 6\alpha \delta\dot\Phi \dot\varphi + 3 \alpha\omega_0 \delta\dot\Phi H + 6\alpha^2 \omega_0 \delta\dot\Phi \varphi 	\\
		&\quad \, + \Big( 1 + \frac{3\alpha^4 \omega_0^2}{4 \xi^2 k^2} \Big) \delta\Phi^2 - \frac{1}{3} \alpha^2 \omega_0^2 k^2 H^2 - k^2 \Big( \frac43 H + 6\alpha\varphi \Big) \delta\Phi.
	\end{aligned}
\end{equation}
However, the second constraint equation~\eqref{CEqpsh} does not involve $ \psi_h $ itself, which indicates that $ \psi_h $ is actually a Lagrange multiplier and that its value cannot be determined from the constraint equation. We therefore keep $ \psi_h $ in the action and proceed to a Hamiltonian analysis.

\subsection{Hamiltonian analysis}

\par We are now ready to carry out the Hamiltonian analysis based on the second-order effective action~\eqref{SoAS1}, which will finally reveal the existence of a ghost instability.

\par For notational convenience, we denote the four scalar variables in action~\eqref{SoAS1} by
\begin{equation}
	q_1 \equiv \delta\Phi, \qquad q_2 \equiv H, \qquad q_3 \equiv \varphi, \qquad q_4 \equiv \psi_h.
\end{equation}
Their conjugate momenta are introduced as follows:
\begin{align}	
	p_1 &= \frac{\partial \mathcal L_\mathrm{S} }{\partial \dot q_1} 
	= -\frac{\alpha}{4\xi^2} \left(  \frac{\alpha}{4\xi^2k^2} \dot q_1 + \frac{1}{3\alpha} \dot q_2 + \dot q_3 + \frac12 \omega_0 q_2 + \alpha\omega_0 q_3 +  \omega_0 q_4  \right), \label{ConjMom1}	\\ 
	p_2 &= \frac{\partial \mathcal L_\mathrm{S} }{\partial \dot q_2} = -\frac{1}{12\xi^2} ( \dot q_1 + \alpha^2 \omega_0^2 \dot q_2  ), \\
	p_3 &= \frac{\partial \mathcal L_\mathrm{S} }{\partial \dot q_3} = -\frac{\alpha}{4\xi^2} \dot q_1,	\label{ConjMom3} \\
	p_4 &= \frac{\partial \mathcal L_\mathrm{S} }{\partial \dot q_4} = 0.	\label{ConjMom4}
\end{align}
The phase space is then parametrized by four pairs of canonical variables,
\begin{equation}	\label{CanVar}
	\{ q_a, p_b \} = \delta_{ab}, \qquad a,b = 1,2,3,4. 
\end{equation}

\par To construct the Hamiltonian, which is a functional of only canonical variables, we first need to express the velocities in terms of these variables. From Eq.~\eqref{ConjMom1}-\eqref{ConjMom3}, it is easy to solve for the velocities,
\begin{align}	
	\dot q_1 &= -\frac{4\xi^2}{\alpha} p_3, \label{dqReplace1}	\\
	\dot q_2 &= \frac{4\xi^2}{\alpha^3 \omega_0^2} (p_3 - 3\alpha p_2 )  , 	\\
	\dot q_3 &= -\frac{4\xi^2}{\alpha} p_1 + \frac{4\xi^2}{\alpha^3 \omega_0^2} p_2 + \left( \frac{1}{k^2} - \frac{4\xi^2}{3\alpha^4 \omega_0^2} \right) p_3 - \frac12 \omega_0 q_2 - \alpha\omega_0 q_3 - \omega_0 q_4. 	\label{dqReplace3}
\end{align}
Note that Eq.~\eqref{ConjMom4} does not determine the velocity $\dot q_4$ but instead yields a primary constraint,
\begin{equation}	\label{PriCon}
	\chi_1 \equiv p_4 \approx 0,
\end{equation}
where the weak equality $ \approx $ denotes equality on the constraint surface. We thus arrive at the canonical Hamiltonian,
\begin{equation}	\label{HamCan}
	\begin{aligned}
		\mathcal H_\mathrm{C} 
		&\equiv \sum_{a=1}^4 p_a \dot q_a - \mathcal L_\mathrm{S} 	\\
		&=  \mathcal H_0 (p_1, p_2, p_3; q_1, q_2, q_3 ) - q_4 \left(\omega_0 p_3 + \frac{3 \alpha^2 \omega_0^2 - 2k^2 }{12\xi^2} q_1 - \frac{\alpha^2 \omega_0^2 k^2 }{6\xi^2} q_2 \right),	\\
		\mathcal H_0
		&=  - \frac{6\xi^2}{\alpha^2 \omega_0^2} p_2^2 + \left( \frac{1}{2k^2} - \frac{2\xi^2}{3\alpha^4 \omega_0^2} \right) p_3^2 - \frac{4\xi^2}{\alpha} p_1p_3 + \frac{4\xi^2}{\alpha^3 \omega_0^2} p_2p_3 - \frac{\omega_0}{2} p_3 q_2 - \alpha \omega_0 p_3 q_3	\\
		&\quad\ \! + \left( \frac{1}{24\xi^2} + \frac{\alpha^4 \omega_0^2}{32\xi^4 k^2} \right) q_1^2  - \frac{\alpha^2 \omega_0^2 k^2}{72\xi^2} q_2^2 - \frac{k^2}{18\xi^2} q_1 q_2 - \frac{\alpha k^2}{4\xi^2} q_1 q_3,
	\end{aligned}
\end{equation}
where the velocities $ \dot q_1, \dot q_2, \dot q_3  $ are eliminated by Eqs.~\eqref{dqReplace1}-~\eqref{dqReplace3}, while the velocity $ \dot q_4 $ cancels out under the primary constraint~\eqref{PriCon}. Taking the primary constraint~\eqref{PriCon} into account, one obtains the total Hamiltonian,
\begin{equation}	\label{HamTot}
	\mathcal H_\mathrm{T} = \mathcal H_\mathrm{C} + \lambda_1 \chi_1,
\end{equation}
where $\lambda_1$ is a Lagrange multiplier.

\par The preservation of the primary constraint $\chi_1$ under time evolution is examined by evaluating its Poisson bracket with the total Hamiltonian $ \mathcal H_\mathrm{T} $, 
\begin{equation}	
	\chi_2 \equiv \dot \chi_1 = \{ \chi_1, \mathcal H_\mathrm{T} \} = \omega_0 p_3 + \frac{1}{12\xi^2} \left[ (3\alpha^2\omega_0^2 - 2k^2) q_1 - 2\alpha^2\omega_0^2 k^2 q_2 \right] \approx 0,
\end{equation}
which provides a secondary constraint that will latter eliminate the expression multiplied by $q_4$ in the canonical Hamiltonian~\eqref{HamCan}. Following Dirac's procedure~\cite{Dirac:1950pj}, we obtain a further secondary constraint from the preservation of $\chi_2$,
\begin{equation}
	\chi_3 \equiv \dot\chi_2 = \{ \chi_2, \mathcal H_\mathrm{T} \} = 2k^2 p_2 + \frac{\alpha\omega_0 k^2}{4\xi^2} q_1 \approx 0.
\end{equation} 
Then the preservation of $\chi_3$ yields yet another secondary constraint,
\begin{equation}
	\chi_4 \equiv \dot\chi_3 = \{ \chi_3, \mathcal H_\mathrm{T} \} = \frac{1}{18\xi^2} \left[ 2q_1 + \alpha^2\omega_0^2 ( q_2 - 6 q_4 ) \right] \approx 0.
\end{equation}
At this stage, a direct calculation shows that the preservation of $\chi_4$ yields a condition that explicitly involves the Lagrange multiplier $\lambda_1$. This condition serves to fix $\lambda_1$, rather than generating a new constraint. The Dirac algorithm therefore terminates, leaving us with four constraints $\chi_1$, $\chi_2$, $\chi_3$, $\chi_4$.

\par With the four constraints at hand, we now examine whether they are first-class or second-class types. The Poisson brackets among these constraints define a $4 \times 4$ matrix, whose determinant is found nonvanishing:
\begin{equation}
	\mathbf M_{ab} \equiv \{ \chi_a, \chi_b \} = \frac{\alpha^2 \omega_0^2 k^4}{3\xi^2}
	\begin{pmatrix}
	\ 	0 \      & \ 0 \      & 0  	& 1      \\
		0      & 0      & -1  	& 0 	\\
		0      & 1      & 0 	& -k^2/3 	\\
		-1     & 0      & k^2/3 	& 0 	\\
	\end{pmatrix},
	\ 
	\det \,  \mathbf M_{ab} = \left( \frac{\alpha^2 \omega_0^2 k^4}{3\xi^2} \right)^4 \neq 0.
\end{equation}
This shows that all constraints $\chi_a$ are second-class. The absence of any first-class constraint indicates that there is no residual gauge freedom~\cite{Dirac:1950pj}.

\par Solving the constraints $ \chi_a \approx 0 $ yields
\begin{equation}	\label{qppq}
	\begin{aligned}
		q_2 &= \frac{6\xi^2}{\alpha^2\omega_0 k^2} p_3 + \left( \frac{3}{2k^2} - \frac{1}{\alpha^2\omega_0^2} \right) q_1 , \qquad
		p_2 = - \frac{\alpha\omega_0}{8\xi^2} q_1,	\\
		q_4 &= \frac{\xi^2}{\alpha^2\omega_0 k^2} p_3 + \left( \frac{1}{4k^2} + \frac{1}{6\alpha^2\omega_0^2} \right) q_1, \qquad
		p_4 = 0,
	\end{aligned}
\end{equation}
which eliminate four of the eight initial canonical variables. The remaining four independent variables are $ \{ q_1, q_3, p_1, p_3 \} $. They represent two scalar degrees of freedom, in agreement with the result in Sec.~\ref{Subsec. ScalarModes}.

\par Substituting the solutions~\eqref{qppq} into the canonical Hamiltonian~\eqref{HamCan}, we obtain the physical Hamiltonian,
\begin{equation}	\label{HamPhy}
	\begin{aligned}
		\mathcal H_{\mathrm{Phy}}
		&= \mathcal C p_3^2 - \frac{4\xi^2}{\alpha} p_1 p_3
		- \left( \frac{1}{6\alpha^2\omega_0} + \frac{\omega_0}{k^2} \right) p_3 q_1 - \alpha\omega_0 p_3 q_3	\\
		&\quad\ +\frac{1}{32\xi^2} \left[ \frac{4k^2}{3\alpha^2\omega_0^2} + \frac{\alpha^2\omega_0^2}{\xi^2 k^2} ( \alpha^2 - \xi^2 ) - 3 \right] q_1^2
		- \frac{\alpha k^2}{4\xi^2} q_1 q_3,
	\end{aligned}
\end{equation}
with a generally nonvanishing coefficient $ \mathcal C \equiv  \dfrac{3\alpha^2 (\alpha^2 - 7\xi^2) \omega_0^2 - 4\xi^2k^2 }{6\alpha^4\omega_0^2 k^2} $. To examine the presence of the ghost instability, we focus on the kinetic terms and rewrite it in a diagonalized form,
\begin{equation}	\label{HamKin}
	\begin{aligned}
		\mathcal H_{\mathrm{Kin}}
		&\equiv \mathcal C p_3^2 - \frac{4\xi^2}{\alpha} p_1 p_3 = \mathcal C \left( p_3 - \frac{2\xi^2}{\alpha \mathcal C}  p_1 \right)^2 - \frac{4\xi^4}{\alpha^2 \mathcal C} p_1^2.
	\end{aligned}
\end{equation}
It is remarkable that the two kinetic terms carry opposite signs, and therefore the kinetic sector is indefinite. Consequently, the Hamiltonian is unbounded from below along specific directions in phase space, confirming the presence of a ghost degree of freedom. This unboundedness can be shown, for example, by fixing $ q_1 = q_3 = 0 $, where one finds $ \mathcal H_{\mathrm{Phy}} \to - \infty $ as $ p_1 \to + \infty $ along $ p_3 = \frac{2\xi^2}{\alpha\mathcal C} p_1 $ for $\mathcal C>0$, or as $ p_3 \to + \infty $ along $ p_1 = 0 $ for $\mathcal C<0$. Alternatively, a direct calculation reveals that the Lagrangian corresponding to the Hamiltonian~\eqref{HamPhy} also features two kinetic terms with opposite signs, which is consistent with the presence of a ghost degree of freedom.

\section{Conclusion}	\label{Sec.Concl}

\par In this work, we have investigated the GW polarization modes and performed the stability analysis within the framework of Weyl geometry gravity. We began with a concise overview of Weyl geometry gravity, presenting the construction of the action and field equations. Subsequently, we carried out a linear perturbation analysis on a homogeneous and isotropic Minkowski background with a constant timelike Weyl gauge field, and constructed a set of gauge invariants, which form the basis for analyzing the polarization modes of GWs and the stability of Weyl geometry gravity.

\par After briefly introducing the polarization modes, we derived the perturbation equations in terms of the gauge invariants, which allow for a systematic analysis of the polarization modes. Our results show that two standard tensor modes propagate at the speed of light, while no vector mode appears, despite the presence of two dynamical vector degrees of freedom. The most distinctive features arise in the scalar sector, where a mixture mode of breathing and longitudinal modes is induced by a single scalar degree of freedom. In addition, there exists another dynamical scalar degree of freedom that does not contribute to the polarization modes.

\par We further discussed the scalar modes and their detectability. The propagating solutions of this mode exhibit superluminal propagation and can undergo intrinsic amplitude decay due to the nonvanishing background vector field. Avoiding a growing homogeneous contribution from the other scalar degree of freedom additionally requires an appropriate choice of initial conditions. Using GW150914 as a representative source, we performed a rough estimate of its observational prospects. We found that, within the decaying propagating branch, the requirement of detecting the scalar modes before their amplitude decay to $ 1/e $ constrains the model parameters such that the scalar and tensor GW signals must arrive nearly simultaneously; otherwise, the scalar-mode amplitude becomes too small to be observable. Consequently, tests of extra polarization modes in Weyl geometry gravity alone do not require a very long searching window in observation time. In practice, however, longer observation windows remain necessary when multiple gravity theories are considered~\cite{Schumacher:2023jxq, Schumacher:2025plq}.

\par Furthermore, our stability analysis reveals a nontrivial structure of the theory. While the tensor and vector sectors are free from both ghost and Laplacian instabilities, the scalar sector exhibits a fundamental inconsistency. A Hamiltonian analysis identifies two dynamical degrees of freedom in the scalar sector, one of which is a ghost, rendering the Hamiltonian unbounded from below. These results suggest that the scalar sector may require further theoretical refinement or additional constraints in order to ensure the long-term stability of Weyl geometry gravity.

\par In summary, this study provides a theoretical foundation for testing Weyl geometry gravity using future GW detectors, such as LISA~\cite{LISA:2017pwj}, Taiji~\cite{Hu:2017mde}, and TianQin~\cite{TianQin:2015yph}. The unique propagation characteristics of the scalar modes offer a possible avenue for multi-messenger observations to probe the geometric structure of spacetime beyond the Riemannian manifold.

\section*{Acknowledgments}
We are sincerely grateful to the anonymous referee for the insightful and constructive comments, which have greatly helped us improve the manuscript. This work is supported in part by the National Key Research and Development Program of China (Grant No. 2021YFC2203003), the National Natural Science Foundation of China (Grants No. 12475056, No. 123B2074, and No. 12247101), Gansu Province's Top Leading Talent Support Plan, the Fundamental Research Funds for the Central Universities (Grant No. lzujbky-2025-jdzx07), the Natural Science Foundation of Gansu Province (No. 22JR5RA389 and No. 25JRRA799), and the 111 Project (Grant No. B20063).

\appendix

\normalem
\bibliography{myref.bib}

@article{LIGOScientific:2016aoc,
	author = "Abbott, B. P. and others",
	collaboration = "LIGO Scientific, Virgo",
	title = "{Observation of Gravitational Waves from a Binary Black Hole Merger}",
	eprint = "1602.03837",
	archivePrefix = "arXiv",
	primaryClass = "gr-qc",
	reportNumber = "LIGO-P150914",
	doi = "10.1103/PhysRevLett.116.061102",
	journal = "Phys. Rev. Lett.",
	volume = "116",
	number = "6",
	pages = "061102",
	year = "2016"
}

@article{LIGOScientific:2016emj,
	author = "Abbott, B. P. and others",
	collaboration = "LIGO Scientific, Virgo",
	title = "{GW150914: The Advanced LIGO Detectors in the Era of First Discoveries}",
	eprint = "1602.03838",
	archivePrefix = "arXiv",
	primaryClass = "gr-qc",
	reportNumber = "LIGO-P1500237",
	doi = "10.1103/PhysRevLett.116.131103",
	journal = "Phys. Rev. Lett.",
	volume = "116",
	number = "13",
	pages = "131103",
	year = "2016"
}

@article{Abbott1,
	author = "Abbott, B. P. and others",
	collaboration = "LIGO Scientific, Virgo",
	title = "{GW151226: Observation of Gravitational Waves from a 22-Solar-Mass Binary Black Hole Coalescence}",
	eprint = "1606.04855",
	archivePrefix = "arXiv",
	primaryClass = "gr-qc",
	reportNumber = "LIGO-P151226",
	doi = "10.1103/PhysRevLett.116.241103",
	journal = "Phys. Rev. Lett.",
	volume = "116",
	number = "24",
	pages = "241103",
	year = "2016"
}

@article{Abbott2,
	author = "Abbott, B. P. and others",
	collaboration = "LIGO Scientific, Virgo",
	title = "{Binary Black Hole Mergers in the first Advanced LIGO Observing Run}",
	eprint = "1606.04856",
	archivePrefix = "arXiv",
	primaryClass = "gr-qc",
	reportNumber = "LIGO-P1600088",
	doi = "10.1103/PhysRevX.6.041015",
	journal = "Phys. Rev. X",
	volume = "6",
	number = "4",
	pages = "041015",
	year = "2016",
	note = "[Erratum: Phys.Rev.X 8, 039903 (2018)]"
}

@article{cai2022spacetime,
	title={Spacetime singularities and cosmic censorship conjectures},
	author={Cai, Rong-Gen and Cao, Li-Ming and Li, Li and Yang, Run-Qiu},
	journal = "Sci. China Phys. Mech. Astron.",
	volume={52},
	number={11},
	pages={110401},
	year={2022}
}

@article{smith1936mass,
	author = "Smith, Sinclair",
	title = "{The Mass of the Virgo Cluster}",
	doi = "10.1086/143697",
	journal = "Astrophys. J.",
	volume = "83",
	pages = "23--30",
	year = "1936"
}

@article{Zwicky:1937zza,
	author = "Zwicky, F.",
	title = "{On the Masses of Nebulae and of Clusters of Nebulae}",
	doi = "10.1086/143864",
	journal = "Astrophys. J.",
	volume = "86",
	pages = "217--246",
	year = "1937"
}

@article{Peebles:2002gy,
	author = "Peebles, P. J. E. and Ratra, Bharat",
	editor = "Hsu, Jong-Ping and Fine, D.",
	title = "{The Cosmological Constant and Dark Energy}",
	eprint = "astro-ph/0207347",
	archivePrefix = "arXiv",
	reportNumber = "KSUPT-02-3",
	doi = "10.1103/RevModPhys.75.559",
	journal = "Rev. Mod. Phys.",
	volume = "75",
	pages = "559--606",
	year = "2003"
}

@article{tHooft:1974toh,
	author = "'t Hooft, Gerard and Veltman, M. J. G.",
	title = "{One loop divergencies in the theory of gravitation}",
	journal = "Ann. Inst. H. Poincare Phys. Theor. A",
	volume = "20",
	pages = "69--94",
	year = "1974"
}

@article{Goroff:1985th,
	author = "Goroff, Marc H. and Sagnotti, Augusto",
	title = "{The Ultraviolet Behavior of Einstein Gravity}",
	reportNumber = "CALT-68-1289, LBL-19995, UCB-PTH-85-34",
	doi = "10.1016/0550-3213(86)90193-8",
	journal = "Nucl. Phys. B",
	volume = "266",
	pages = "709--736",
	year = "1986"
}

@article{Han:2005km,
	author = "Han, Muxin and Huang, Weiming and Ma, Yongge",
	title = "{Fundamental structure of loop quantum gravity}",
	eprint = "gr-qc/0509064",
	archivePrefix = "arXiv",
	doi = "10.1142/S0218271807010894",
	journal = "Int. J. Mod. Phys. D",
	volume = "16",
	pages = "1397--1474",
	year = "2007"
}

@article{clifton2012modified,
	author = "Clifton, Timothy and Ferreira, Pedro G. and Padilla, Antonio and Skordis, Constantinos",
	title = "{Modified Gravity and Cosmology}",
	eprint = "1106.2476",
	archivePrefix = "arXiv",
	primaryClass = "astro-ph.CO",
	doi = "10.1016/j.physrep.2012.01.001",
	journal = "Phys. Rept.",
	volume = "513",
	pages = "1--189",
	year = "2012"
}

@article{Weyl:1918ib,
	author = "Weyl, H.",
	title = "{Gravitation and electricity}",
	journal = "Sitzungsber. Preuss. Akad. Wiss. Berlin (Math. Phys. )",
	volume = "1918",
	pages = "465",
	year = "1918"
}

@article{Lobo:2018zrz,
	author = "Lobo, I. P. and Romero, C.",
	title = "{Experimental constraints on the second clock effect}",
	eprint = "1807.07188",
	archivePrefix = "arXiv",
	primaryClass = "gr-qc",
	doi = "10.1016/j.physletb.2018.07.019",
	journal = "Phys. Lett. B",
	volume = "783",
	pages = "306--310",
	year = "2018"
}

@article{Scholz:2017pfo,
	author = "Scholz, Erhard",
	title = "{The unexpected resurgence of Weyl geometry in late 20-th century physics}",
	eprint = "1703.03187",
	archivePrefix = "arXiv",
	primaryClass = "math.HO",
	doi = "10.1007/978-1-4939-7708-6_11",
	journal = "Einstein Stud.",
	volume = "14",
	pages = "261--360",
	year = "2018"
}

@article{Ghilencea:2018thl,
	author = "Ghilencea, D. M. and Lee, Hyun Min",
	title = "{Weyl gauge symmetry and its spontaneous breaking in the standard model and inflation}",
	eprint = "1809.09174",
	archivePrefix = "arXiv",
	primaryClass = "hep-th",
	doi = "10.1103/PhysRevD.99.115007",
	journal = "Phys. Rev. D",
	volume = "99",
	number = "11",
	pages = "115007",
	year = "2019"
}

@article{Ghilencea:2018dqd,
	author = "Ghilencea, D. M.",
	title = "{Spontaneous breaking of Weyl quadratic gravity to Einstein action and Higgs potential}",
	eprint = "1812.08613",
	archivePrefix = "arXiv",
	primaryClass = "hep-th",
	doi = "10.1007/JHEP03(2019)049",
	journal = "JHEP",
	volume = "03",
	pages = "049",
	year = "2019"
}

@article{Ghilencea:2019jux,
	author = "Ghilencea, D. M.",
	title = "{Stueckelberg breaking of Weyl conformal geometry and applications to gravity}",
	eprint = "1904.06596",
	archivePrefix = "arXiv",
	primaryClass = "hep-th",
	doi = "10.1103/PhysRevD.101.045010",
	journal = "Phys. Rev. D",
	volume = "101",
	number = "4",
	pages = "045010",
	year = "2020"
}

@article{Ghilencea:2019rqj,
	author = "Ghilencea, D. M.",
	title = "{Weyl R$^{2}$ inflation with an emergent Planck scale}",
	eprint = "1906.11572",
	archivePrefix = "arXiv",
	primaryClass = "gr-qc",
	doi = "10.1007/JHEP10(2019)209",
	journal = "JHEP",
	volume = "10",
	pages = "209",
	year = "2019"
}

@article{Ghilencea:2021jjl,
	author = "Ghilencea, D. M. and Harko, T.",
	title = "{Cosmological evolution in Weyl conformal geometry}",
	eprint = "2110.07056",
	archivePrefix = "arXiv",
	primaryClass = "gr-qc",
	month = "10",
	year = "2021"
}

@article{Burikham:2023bil,
	author = "Burikham, Piyabut and Harko, Tiberiu and Pimsamarn, Kulapant and Shahidi, Shahab",
	title = "{Dark matter as a Weyl geometric effect}",
	eprint = "2302.08289",
	archivePrefix = "arXiv",
	primaryClass = "gr-qc",
	doi = "10.1103/PhysRevD.107.064008",
	journal = "Phys. Rev. D",
	volume = "107",
	number = "6",
	pages = "064008",
	year = "2023"
}

@article{Haghani:2023nrm,
	author = "Haghani, Zahra and Harko, Tiberiu",
	title = "{Compact stellar structures in Weyl geometric gravity}",
	eprint = "2303.10339",
	archivePrefix = "arXiv",
	primaryClass = "gr-qc",
	doi = "10.1103/PhysRevD.107.064068",
	journal = "Phys. Rev. D",
	volume = "107",
	number = "6",
	pages = "064068",
	year = "2023"
}

@article{Khodadi:2025gtq,
	author = "Khodadi, Mohsen and Harko, Tiberiu",
	title = "{Weyl geometric gravity black holes in light of the Solar System tests}",
	eprint = "2509.15838",
	archivePrefix = "arXiv",
	primaryClass = "gr-qc",
	doi = "10.1140/epjc/s10052-025-14982-5",
	journal = "Eur. Phys. J. C",
	volume = "85",
	number = "11",
	pages = "1325",
	year = "2025"
}

@article{Oancea:2023ylb,
	author = "Oancea, Marius A. and Harko, Tiberiu",
	title = "{Weyl geometric effects on the propagation of light in gravitational fields}",
	eprint = "2305.01313",
	archivePrefix = "arXiv",
	primaryClass = "gr-qc",
	doi = "10.1103/PhysRevD.109.064020",
	journal = "Phys. Rev. D",
	volume = "109",
	number = "6",
	pages = "064020",
	year = "2024"
}

@article{Visa:2024fii,
	author = "Visa, Daria-Ioana and Harko, Tiberiu and Shahidi, Shahab",
	title = "{Mimetic Weyl geometric gravity}",
	eprint = "2410.22787",
	archivePrefix = "arXiv",
	primaryClass = "gr-qc",
	doi = "10.1016/j.dark.2024.101720",
	journal = "Phys. Dark Univ.",
	volume = "46",
	pages = "101720",
	year = "2024"
}

@article{Lee:2024rjw,
	author = "Lee, Hyun Min",
	title = "{Pole Inflation from Broken Noncompact Isometry in Weyl Gravity}",
	eprint = "2411.16944",
	archivePrefix = "arXiv",
	primaryClass = "hep-ph",
	doi = "10.1103/skhx-yc43",
	journal = "Phys. Rev. Lett.",
	volume = "134",
	number = "21",
	pages = "211001",
	year = "2025"
}

@article{Sanomiya:2020svg,
	author = "Sanomiya, T. A. T. and Lobo, I. P. and Formiga, J. B. and Dahia, F. and Romero, C.",
	editor = "Mostepanenko, V. M. and Velichko, E. N.",
	title = "{Invariant approach to Weyl{\textquoteright}s unified field theory}",
	eprint = "2002.00285",
	archivePrefix = "arXiv",
	primaryClass = "gr-qc",
	doi = "10.1142/S0217751X20400059",
	journal = "Phys. Rev. D",
	volume = "102",
	pages = "124031",
	year = "2020"
}

@article{Duarte:2024zjb,
	author = "Duarte, Mauro and Dahia, Fabio and Romero, Carlos",
	title = "{On the Propagation of Gravitational Waves in the Weyl Invariant Theory of Gravity}",
	doi = "10.3390/universe10090361",
	journal = "Universe",
	volume = "10",
	number = "9",
	pages = "361",
	year = "2024"
}

@article{Khodadi:2026zoi,
	author = "Khodadi, Mohsen and Saridakis, Emmanuel N.",
	title = "{Atomic clocks and gravitational waves as probes of non-metricity}",
	eprint = "2601.19407",
	archivePrefix = "arXiv",
	primaryClass = "gr-qc",
	month = "1",
	year = "2026"
}

@article{Eardley:1973zuo,
	author = "Eardley, D. M. and Lee, D. L. and Lightman, A. P.",
	title = "{Gravitational-wave observations as a tool for testing relativistic gravity}",
	doi = "10.1103/PhysRevD.8.3308",
	journal = "Phys. Rev. D",
	volume = "8",
	pages = "3308--3321",
	year = "1973"
}

@article{Sotiriou:2008rp,
	author = "Sotiriou, Thomas P. and Faraoni, Valerio",
	title = "{f(R) Theories Of Gravity}",
	eprint = "0805.1726",
	archivePrefix = "arXiv",
	primaryClass = "gr-qc",
	doi = "10.1103/RevModPhys.82.451",
	journal = "Rev. Mod. Phys.",
	volume = "82",
	pages = "451--497",
	year = "2010"
}

@article{DeFelice:2010aj,
	author = "De Felice, Antonio and Tsujikawa, Shinji",
	title = "{f(R) theories}",
	eprint = "1002.4928",
	archivePrefix = "arXiv",
	primaryClass = "gr-qc",
	doi = "10.12942/lrr-2010-3",
	journal = "Living Rev. Rel.",
	volume = "13",
	pages = "3",
	year = "2010"
}

@article{Berezin:2021iof,
	author = "Berezin, Victor A. and Dokuchaev, Vyacheslav I. and Eroshenko, Yu. N. and Eroshenko, Yury N. and Smirnov, Alexey L.",
	title = "{On the cosmological solutions in Weyl geometry}",
	eprint = "2107.06160",
	archivePrefix = "arXiv",
	primaryClass = "gr-qc",
	doi = "10.1088/1475-7516/2021/11/053",
	journal = "JCAP",
	volume = "11",
	number = "11",
	pages = "053",
	year = "2021"
}

@article{Berezin:2022odj,
	author = "Berezin, V. A. and Dokuchaev, V. I.",
	title = "{Weyl cosmology}",
	eprint = "2203.04257",
	archivePrefix = "arXiv",
	primaryClass = "gr-qc",
	doi = "10.1142/S0217751X22430059",
	journal = "Int. J. Mod. Phys. A",
	volume = "37",
	number = "20n21",
	pages = "2243005",
	year = "2022"
}

@article{Berezin:2022phu,
	author = "Berezin, V. A. and Dokuchaev, V. I.",
	title = "{Cosmological particle creation in Weyl geometry}",
	eprint = "2207.00057",
	archivePrefix = "arXiv",
	primaryClass = "gr-qc",
	doi = "10.1088/1361-6382/aca57e",
	journal = "Class. Quant. Grav.",
	volume = "40",
	number = "1",
	pages = "015006",
	year = "2023"
}

@article{Liang:2017ahj,
	author = "Liang, Dicong and Gong, Yungui and Hou, Shaoqi and Liu, Yunqi",
	title = "{Polarizations of gravitational waves in $f(R)$ gravity}",
	eprint = "1701.05998",
	archivePrefix = "arXiv",
	primaryClass = "gr-qc",
	doi = "10.1103/PhysRevD.95.104034",
	journal = "Phys. Rev. D",
	volume = "95",
	number = "10",
	pages = "104034",
	year = "2017"
}

@article{Hou:2017bqj,
	author = "Hou, Shaoqi and Gong, Yungui and Liu, Yunqi",
	title = "{Polarizations of Gravitational Waves in Horndeski Theory}",
	eprint = "1704.01899",
	archivePrefix = "arXiv",
	primaryClass = "gr-qc",
	doi = "10.1140/epjc/s10052-018-5869-y",
	journal = "Eur. Phys. J. C",
	volume = "78",
	number = "5",
	pages = "378",
	year = "2018"
}

@article{Gong:2018cgj,
	author = "Gong, Yungui and Hou, Shaoqi and Liang, Dicong and Papantonopoulos, Eleftherios",
	title = "{Gravitational waves in Einstein-{\ae}ther and generalized TeVeS theory after GW170817}",
	eprint = "1801.03382",
	archivePrefix = "arXiv",
	primaryClass = "gr-qc",
	doi = "10.1103/PhysRevD.97.084040",
	journal = "Phys. Rev. D",
	volume = "97",
	number = "8",
	pages = "084040",
	year = "2018"
}

@article{Wagle:2019mdq,
	author = "Wagle, Pratik and Saffer, Alexander and Yunes, Nicolas",
	title = "{Polarization modes of gravitational waves in Quadratic Gravity}",
	eprint = "1910.04800",
	archivePrefix = "arXiv",
	primaryClass = "gr-qc",
	reportNumber = "Phys. Rev. D 100, 124007",
	doi = "10.1103/PhysRevD.100.124007",
	journal = "Phys. Rev. D",
	volume = "100",
	number = "12",
	pages = "124007",
	year = "2019"
}

@article{Dong:2021jtd,
	author = "Dong, Yu-Qi and Liu, Yu-Xiao",
	title = "{Polarization modes of gravitational waves in Palatini-Horndeski theory}",
	eprint = "2111.07352",
	archivePrefix = "arXiv",
	primaryClass = "gr-qc",
	doi = "10.1103/PhysRevD.105.064035",
	journal = "Phys. Rev. D",
	volume = "105",
	number = "6",
	pages = "064035",
	year = "2022"
}

@article{Liang:2022hxd,
	author = "Liang, Dicong and Xu, Rui and Lu, Xuchen and Shao, Lijing",
	title = "{Polarizations of gravitational waves in the bumblebee gravity model}",
	eprint = "2207.14423",
	archivePrefix = "arXiv",
	primaryClass = "gr-qc",
	doi = "10.1103/PhysRevD.106.124019",
	journal = "Phys. Rev. D",
	volume = "106",
	number = "12",
	pages = "124019",
	year = "2022"
}

@article{Dong:2023xyb,
	author = "Dong, Yu-Qi and Liu, Yu-Qiang and Liu, Yu-Xiao",
	title = "{Polarization modes of gravitational waves in generalized Proca theory}",
	eprint = "2305.12516",
	archivePrefix = "arXiv",
	primaryClass = "gr-qc",
	doi = "10.1103/PhysRevD.109.024014",
	journal = "Phys. Rev. D",
	volume = "109",
	number = "2",
	pages = "024014",
	year = "2024"
}

@article{Dong:2023bgt,
	author = "Dong, Yu-Qi and Liu, Yu-Qiang and Liu, Yu-Xiao",
	title = "{Polarization modes of gravitational waves in general modified gravity: General metric theory and general scalar-tensor theory}",
	eprint = "2310.11336",
	archivePrefix = "arXiv",
	primaryClass = "gr-qc",
	doi = "10.1103/PhysRevD.109.044013",
	journal = "Phys. Rev. D",
	volume = "109",
	number = "4",
	pages = "044013",
	year = "2024"
}

@article{Lai:2024fza,
	author = "Lai, Xiao-Bin and Dong, Yu-Qi and Liu, Yu-Qiang and Liu, Yu-Xiao",
	title = "{Polarization modes of gravitational waves in general Einstein-vector theory}",
	eprint = "2405.20577",
	archivePrefix = "arXiv",
	primaryClass = "gr-qc",
	doi = "10.1103/PhysRevD.110.064073",
	journal = "Phys. Rev. D",
	volume = "110",
	number = "6",
	pages = "064073",
	year = "2024"
}

@article{Dong:2024zal,
	author = "Dong, Yu-Qi and Lai, Xiao-Bin and Liu, Yu-Qiang and Liu, Yu-Xiao",
	title = "{Gravitational-wave effects in the most general vector{\textendash}tensor theory}",
	eprint = "2409.11838",
	archivePrefix = "arXiv",
	primaryClass = "gr-qc",
	doi = "10.1140/epjc/s10052-025-14378-5",
	journal = "Eur. Phys. J. C",
	volume = "85",
	number = "6",
	pages = "645",
	year = "2025"
}

@article{Fan:2024pex,
	author = "Fan, Yu-Zhi and Lai, Xiao-Bin and Dong, Yu-Qi and Liu, Yu-Xiao",
	title = "{Polarization modes of gravitational waves in scalar-tensor-Rastall theory}",
	eprint = "2409.18503",
	archivePrefix = "arXiv",
	primaryClass = "gr-qc",
	doi = "10.1140/epjc/s10052-025-13778-x",
	journal = "Eur. Phys. J. C",
	volume = "85",
	number = "1",
	pages = "65",
	year = "2025"
}

@article{Lai:2026vhe,
	author = "Lai, Xiao-Bin and Fan, Yu-Zhi and Dong, Yu-Qi and Liu, Yu-Xiao",
	title = "{Cosmological perturbations and gravitational waves in the general Einstein-vector theory}",
	eprint = "2602.12536",
	archivePrefix = "arXiv",
	primaryClass = "gr-qc",
	doi = "10.1103/57hr-5pt9",
	journal = "Phys. Rev. D",
	volume = "113",
	number = "12",
	pages = "124073",
	year = "2026"
}

@article{Dong:2025ddi,
	author = "Dong, Yu-Qi and Lai, Xiao-Bin and Fan, Yu-Zhi and Liu, Yu-Xiao",
	title = "{New gravitational wave polarization modes in the torsionless spacetime}",
	eprint = "2504.09445",
	archivePrefix = "arXiv",
	primaryClass = "gr-qc",
	doi = "10.1140/epjc/s10052-025-15006-y",
	journal = "Eur. Phys. J. C",
	volume = "85",
	number = "11",
	pages = "1249",
	year = "2025"
}

@article{Dong:2025pyz,
	author = "Dong, Yu-Qi and Lai, Xiao-Bin and Fan, Yu-Zhi and Liu, Yu-Xiao",
	title = "{Polarization modes of gravitational waves in general symmetric teleparallel gravity}",
	eprint = "2505.13298",
	archivePrefix = "arXiv",
	primaryClass = "gr-qc",
	month = "5",
	year = "2025"
}

@article{LIGOScientific:2017ous,
	author = "Abbott, Benjamin P. and others",
	collaboration = "LIGO Scientific, Virgo",
	title = "{First search for nontensorial gravitational waves from known pulsars}",
	eprint = "1709.09203",
	archivePrefix = "arXiv",
	primaryClass = "gr-qc",
	reportNumber = "LIGO-P1700009",
	doi = "10.1103/PhysRevLett.120.031104",
	journal = "Phys. Rev. Lett.",
	volume = "120",
	number = "3",
	pages = "031104",
	year = "2018"
}

@article{LIGOScientific:2017ycc,
	author = "Abbott, B. P. and others",
	collaboration = "LIGO Scientific, Virgo",
	title = "{GW170814: A Three-Detector Observation of Gravitational Waves from a Binary Black Hole Coalescence}",
	eprint = "1709.09660",
	archivePrefix = "arXiv",
	primaryClass = "gr-qc",
	doi = "10.1103/PhysRevLett.119.141101",
	journal = "Phys. Rev. Lett.",
	volume = "119",
	number = "14",
	pages = "141101",
	year = "2017"
}

@article{LIGOScientific:2018czr,
	author = "Abbott, Benjamin P. and others",
	collaboration = "LIGO Scientific, Virgo",
	title = "{Search for Tensor, Vector, and Scalar Polarizations in the Stochastic Gravitational-Wave Background}",
	eprint = "1802.10194",
	archivePrefix = "arXiv",
	primaryClass = "gr-qc",
	doi = "10.1103/PhysRevLett.120.201102",
	journal = "Phys. Rev. Lett.",
	volume = "120",
	number = "20",
	pages = "201102",
	year = "2018"
}

@article{LIGOScientific:2018dkp,
	author = "Abbott, B. P. and others",
	collaboration = "LIGO Scientific, Virgo",
	title = "{Tests of General Relativity with GW170817}",
	eprint = "1811.00364",
	archivePrefix = "arXiv",
	primaryClass = "gr-qc",
	reportNumber = "LIGO-P1800059",
	doi = "10.1103/PhysRevLett.123.011102",
	journal = "Phys. Rev. Lett.",
	volume = "123",
	number = "1",
	pages = "011102",
	year = "2019"
}

@article{LIGOScientific:2019fpa,
	author = "Abbott, B. P. and others",
	collaboration = "LIGO Scientific, Virgo",
	title = "{Tests of General Relativity with the Binary Black Hole Signals from the LIGO-Virgo Catalog GWTC-1}",
	eprint = "1903.04467",
	archivePrefix = "arXiv",
	primaryClass = "gr-qc",
	reportNumber = "LIGO-P1800316",
	doi = "10.1103/PhysRevD.100.104036",
	journal = "Phys. Rev. D",
	volume = "100",
	number = "10",
	pages = "104036",
	year = "2019"
}

@article{LIGOScientific:2021sio,
	author = "Abbott, R. and others",
	collaboration = "LIGO Scientific, VIRGO, KAGRA",
	title = "{Tests of General Relativity with GWTC-3}",
	eprint = "2112.06861",
	archivePrefix = "arXiv",
	primaryClass = "gr-qc",
	reportNumber = "LIGO-P2100275",
	doi = "10.1103/PhysRevD.112.084080",
	journal = "Phys. Rev. D",
	volume = "112",
	number = "8",
	pages = "084080",
	year = "2025"
}

@article{Yunes:2013dva,
	author = "Yunes, Nicol{\'a}s and Siemens, Xavier",
	title = "{Gravitational-Wave Tests of General Relativity with Ground-Based Detectors and Pulsar Timing-Arrays}",
	eprint = "1304.3473",
	archivePrefix = "arXiv",
	primaryClass = "gr-qc",
	doi = "10.12942/lrr-2013-9",
	journal = "Living Rev. Rel.",
	volume = "16",
	pages = "9",
	year = "2013"
}

@article{LISACosmologyWorkingGroup:2026zah,
	author = "Akama, Shingo and others",
	collaboration = "LISA Cosmology Working Group",
	title = "{Testing gravitational wave polarizations with LISA}",
	eprint = "2603.03165",
	archivePrefix = "arXiv",
	primaryClass = "astro-ph.CO",
	month = "3",
	year = "2026"
}

@article{Liu:2020mab,
	author = "Liu, Chang and Ruan, Wen-Hong and Guo, Zong-Kuan",
	title = "{Constraining gravitational-wave polarizations with Taiji}",
	eprint = "2006.04413",
	archivePrefix = "arXiv",
	primaryClass = "gr-qc",
	doi = "10.1103/PhysRevD.102.124050",
	journal = "Phys. Rev. D",
	volume = "102",
	number = "12",
	pages = "124050",
	year = "2020"
}

@article{Amalberti:2021kzh,
	author = "Amalberti, Loris and Bartolo, Nicola and Ricciardone, Angelo",
	title = "{Sensitivity of third-generation interferometers to extra polarizations in the stochastic gravitational wave background}",
	eprint = "2105.13197",
	archivePrefix = "arXiv",
	primaryClass = "astro-ph.CO",
	doi = "10.1103/PhysRevD.105.064033",
	journal = "Phys. Rev. D",
	volume = "105",
	number = "6",
	pages = "064033",
	year = "2022"
}

@article{ET:2025xjr,
	author = "Abac, Adrian and others",
	collaboration = "ET",
	title = "{The Science of the Einstein Telescope}",
	eprint = "2503.12263",
	archivePrefix = "arXiv",
	primaryClass = "gr-qc",
	reportNumber = "ET-0036C-25",
	doi = "10.1088/1475-7516/2026/03/081",
	journal = "JCAP",
	volume = "03",
	pages = "081",
	year = "2026"
}

@article{Sazhin:1978myk,
	author = "Sazhin, Mikhail V.",
	title = "{Opportunities for detecting ultralong gravitational waves}",
	journal = "Sov. Astron.",
	volume = "22",
	pages = "36--38",
	year = "1978"
}

@article{Detweiler:1979wn,
	author = "Detweiler, Steven L.",
	title = "{Pulsar timing measurements and the search for gravitational waves}",
	doi = "10.1086/157593",
	journal = "Astrophys. J.",
	volume = "234",
	pages = "1100--1104",
	year = "1979"
}

@article{Chen:2021wdo,
	author = "Chen, Zu-Cheng and Yuan, Chen and Huang, Qing-Guo",
	title = "{Non-tensorial gravitational wave background in NANOGrav 12.5-year data set}",
	eprint = "2101.06869",
	archivePrefix = "arXiv",
	primaryClass = "astro-ph.CO",
	doi = "10.1007/s11433-021-1797-y",
	journal = "Sci. China Phys. Mech. Astron.",
	volume = "64",
	number = "12",
	pages = "12",
	year = "2021"
}

@article{Chen:2023uiz,
	author = "Chen, Zu-Cheng and Wu, Yu-Mei and Bi, Yan-Chen and Huang, Qing-Guo",
	title = "{Search for nontensorial gravitational-wave backgrounds in the NANOGrav 15-year dataset}",
	eprint = "2310.11238",
	archivePrefix = "arXiv",
	primaryClass = "astro-ph.CO",
	doi = "10.1103/PhysRevD.109.084045",
	journal = "Phys. Rev. D",
	volume = "109",
	number = "8",
	pages = "084045",
	year = "2024"
}

@article{LISA:2017pwj,
	author = "Amaro-Seoane, Pau and others",
	collaboration = "LISA",
	title = "{Laser Interferometer Space Antenna}",
	eprint = "1702.00786",
	archivePrefix = "arXiv",
	primaryClass = "astro-ph.IM",
	month = "2",
	year = "2017"
}

@article{Hu:2017mde,
	author = "Hu, Wen-Rui and Wu, Yue-Liang",
	title = "{The Taiji Program in Space for gravitational wave physics and the nature of gravity}",
	doi = "10.1093/nsr/nwx116",
	journal = "Natl. Sci. Rev.",
	volume = "4",
	number = "5",
	pages = "685--686",
	year = "2017"
}

@article{TianQin:2015yph,
	author = "Luo, Jun and others",
	collaboration = "TianQin",
	title = "{TianQin: a space-borne gravitational wave detector}",
	eprint = "1512.02076",
	archivePrefix = "arXiv",
	primaryClass = "astro-ph.IM",
	doi = "10.1088/0264-9381/33/3/035010",
	journal = "Class. Quant. Grav.",
	volume = "33",
	number = "3",
	pages = "035010",
	year = "2016"
}

@article{Tinto:2010hz,
	author = "Tinto, Massimo and da Silva Alves, Marcio Eduardo",
	title = "{LISA Sensitivities to Gravitational Waves from Relativistic Metric Theories of Gravity}",
	eprint = "1010.1302",
	archivePrefix = "arXiv",
	primaryClass = "gr-qc",
	doi = "10.1103/PhysRevD.82.122003",
	journal = "Phys. Rev. D",
	volume = "82",
	pages = "122003",
	year = "2010"
}

@article{LISA:2024hlh,
	author = "Colpi, Monica and others",
	collaboration = "LISA",
	title = "{LISA Definition Study Report}",
	eprint = "2402.07571",
	archivePrefix = "arXiv",
	primaryClass = "astro-ph.CO",
	month = "2",
	year = "2024"
}

@article{Zhang:2019oet,
	author = "Zhang, Chunyu and Gao, Qing and Gong, Yungui and Liang, Dicong and Weinstein, Alan J. and Zhang, Chao",
	title = "{Frequency response of time-delay interferometry for space-based gravitational wave antenna}",
	eprint = "1906.10901",
	archivePrefix = "arXiv",
	primaryClass = "gr-qc",
	doi = "10.1103/PhysRevD.100.064033",
	journal = "Phys. Rev. D",
	volume = "100",
	number = "6",
	pages = "064033",
	year = "2019"
}

@article{Zhang:2020khm,
	author = "Zhang, Chunyu and Gao, Qing and Gong, Yungui and Wang, Bin and Weinstein, Alan J. and Zhang, Chao",
	title = "{Full analytical formulas for frequency response of space-based gravitational wave detectors}",
	eprint = "2003.01441",
	archivePrefix = "arXiv",
	primaryClass = "gr-qc",
	doi = "10.1103/PhysRevD.101.124027",
	journal = "Phys. Rev. D",
	volume = "101",
	number = "12",
	pages = "124027",
	year = "2020"
}

@article{Philippoz:2017ywb,
	author = "Philippoz, L. and Jetzer, P.",
	editor = "Giardini, Domencio and Jetzer, Philippe",
	title = "{Detecting additional polarization modes with LISA}",
	doi = "10.1088/1742-6596/840/1/012057",
	journal = "J. Phys. Conf. Ser.",
	volume = "840",
	number = "1",
	pages = "012057",
	year = "2017"
}

@article{Wang:2021mou,
	author = "Wang, Gang and Han, Wen-Biao",
	title = "{Observing gravitational wave polarizations with the LISA-TAIJI network}",
	eprint = "2101.01991",
	archivePrefix = "arXiv",
	primaryClass = "gr-qc",
	doi = "10.1103/PhysRevD.103.064021",
	journal = "Phys. Rev. D",
	volume = "103",
	number = "6",
	pages = "064021",
	year = "2021"
}

@article{Hu:2022byd,
	author = "Hu, Yu and Wang, Pan-Pan and Tan, Yu-Jie and Shao, Cheng-Gang",
	title = "{Testing the polarization of gravitational wave background with LISA-TianQin network}",
	eprint = "2209.07049",
	archivePrefix = "arXiv",
	primaryClass = "gr-qc",
	month = "9",
	year = "2022"
}

@article{Woodard:2006nt,
	author = "Woodard, Richard P.",
	editor = "Papantonopoulos, Lefteris",
	title = "{Avoiding dark energy with 1/r modifications of gravity}",
	eprint = "astro-ph/0601672",
	archivePrefix = "arXiv",
	reportNumber = "UFIFT-QG-06-02",
	doi = "10.1007/978-3-540-71013-4_14",
	journal = "Lect. Notes Phys.",
	volume = "720",
	pages = "403--433",
	year = "2007"
}

@article{DeFelice:2010gb,
	author = "De Felice, Antonio and Mukohyama, Shinji and Tsujikawa, Shinji",
	title = "{Density perturbations in general modified gravitational theories}",
	eprint = "1006.0281",
	archivePrefix = "arXiv",
	primaryClass = "astro-ph.CO",
	doi = "10.1103/PhysRevD.82.023524",
	journal = "Phys. Rev. D",
	volume = "82",
	pages = "023524",
	year = "2010"
}

@article{Bean:2007nx,
	author = "Bean, Rachel and Flanagan, Eanna E. and Trodden, Mark",
	title = "{The Adiabatic Instability on Cosmology's Dark Side}",
	eprint = "0709.1124",
	archivePrefix = "arXiv",
	primaryClass = "astro-ph",
	doi = "10.1088/1367-2630/10/3/033006",
	journal = "New J. Phys.",
	volume = "10",
	pages = "033006",
	year = "2008"
}

@article{DeFelice:2016yws,
	author = "De Felice, Antonio and Heisenberg, Lavinia and Kase, Ryotaro and Mukohyama, Shinji and Tsujikawa, Shinji and Zhang, Ying-li",
	title = "{Cosmology in generalized Proca theories}",
	eprint = "1603.05806",
	archivePrefix = "arXiv",
	primaryClass = "gr-qc",
	reportNumber = "YITP-16-36, IPMU16-0032",
	doi = "10.1088/1475-7516/2016/06/048",
	journal = "JCAP",
	volume = "06",
	pages = "048",
	year = "2016"
}

@article{Kase:2018aps,
	author = "Kase, Ryotaro and Tsujikawa, Shinji",
	title = "{Dark energy in Horndeski theories after GW170817: A review}",
	eprint = "1809.08735",
	archivePrefix = "arXiv",
	primaryClass = "gr-qc",
	doi = "10.1142/S0218271819420057",
	journal = "Int. J. Mod. Phys. D",
	volume = "28",
	number = "05",
	pages = "1942005",
	year = "2019"
}

@article{Clough:2022ygm,
	author = "Clough, Katy and Helfer, Thomas and Witek, Helvi and Berti, Emanuele",
	title = "{Ghost Instabilities in Self-Interacting Vector Fields: The Problem with Proca Fields}",
	eprint = "2204.10868",
	archivePrefix = "arXiv",
	primaryClass = "gr-qc",
	doi = "10.1103/PhysRevLett.129.151102",
	journal = "Phys. Rev. Lett.",
	volume = "129",
	number = "15",
	pages = "151102",
	year = "2022"
}

@article{vandeBruck:2025aaa,
	author = "van de Bruck, Carsten and Gorji, Mohammad Ali and Nilsson, Nils A. and Pookkillath, Masroor C. and Yamaguchi, Masahide",
	title = "{A no-go theorem in bumblebee vector-tensor cosmology}",
	eprint = "2509.11647",
	archivePrefix = "arXiv",
	primaryClass = "hep-th",
	reportNumber = "CQUeST-2026-0770",
	doi = "10.1088/1475-7516/2026/07/043",
	journal = "JCAP",
	volume = "07",
	pages = "043",
	year = "2026"
}

@article{Lai:2025nyo,
	author = "Lai, Xiao-Bin and Dong, Yu-Qi and Fan, Yu-Zhi and Liu, Yu-Xiao",
	title = "{Stability analysis of cosmological perturbations in the bumblebee model: Parameter constraints and gravitational waves}",
	eprint = "2509.13958",
	archivePrefix = "arXiv",
	primaryClass = "gr-qc",
	doi = "10.1103/q6fk-3lkj",
	journal = "Phys. Rev. D",
	volume = "113",
	number = "4",
	pages = "044003",
	year = "2026"
}

@article{Ostrogradsky:1850fid,
	author = "Ostrogradsky, M.",
	title = "{M{\'e}moires sur les {\'e}quations diff{\'e}rentielles, relatives au probl{\`e}me des isop{\'e}rim{\`e}tres}",
	journal = "Mem. Acad. St. Petersbourg",
	volume = "6",
	number = "4",
	pages = "385--517",
	year = "1850"
}

@article{Noether_1971,
	title={Invariant variation problems},
	volume={1},
	ISSN={1532-2424},
	url={http://dx.doi.org/10.1080/00411457108231446},
	DOI={10.1080/00411457108231446},
	number={3},
	journal={Transport Theory and Statistical Physics},
	publisher={Informa UK Limited},
	author={Noether, Emmy},
	year={1971},
	month=jan, pages={186–207} }

@article{Ruzziconi:2019pzd,
	author = "Ruzziconi, Romain",
	title = "{Asymptotic Symmetries in the Gauge Fixing Approach and the BMS Group}",
	eprint = "1910.08367",
	archivePrefix = "arXiv",
	primaryClass = "hep-th",
	doi = "10.22323/1.384.0003",
	journal = "PoS",
	volume = "Modave2019",
	pages = "003",
	year = "2020"
}

@article{Bardeen:1980kt,
	author = "Bardeen, James M.",
	title = "{Gauge Invariant Cosmological Perturbations}",
	doi = "10.1103/PhysRevD.22.1882",
	journal = "Phys. Rev. D",
	volume = "22",
	pages = "1882--1905",
	year = "1980"
}

@article{Flanagan:2005yc,
	author = "Flanagan, Eanna E. and Hughes, Scott A.",
	title = "{The Basics of gravitational wave theory}",
	eprint = "gr-qc/0501041",
	archivePrefix = "arXiv",
	doi = "10.1088/1367-2630/7/1/204",
	journal = "New J. Phys.",
	volume = "7",
	pages = "204",
	year = "2005"
}

@article{Caprini:2018mtu,
	author = "Caprini, Chiara and Figueroa, Daniel G.",
	title = "{Cosmological Backgrounds of Gravitational Waves}",
	eprint = "1801.04268",
	archivePrefix = "arXiv",
	primaryClass = "astro-ph.CO",
	doi = "10.1088/1361-6382/aac608",
	journal = "Class. Quant. Grav.",
	volume = "35",
	number = "16",
	pages = "163001",
	year = "2018"
}

@article{Alves:2023rxs,
	author = "Alves, M{\'a}rcio E. S.",
	title = "{Testing gravity with gauge-invariant polarization states of gravitational waves: Theory and pulsar timing sensitivity}",
	eprint = "2308.09178",
	archivePrefix = "arXiv",
	primaryClass = "gr-qc",
	doi = "10.1103/PhysRevD.109.104054",
	journal = "Phys. Rev. D",
	volume = "109",
	number = "10",
	pages = "104054",
	year = "2024"
}

@book{Maggiore:2007ulw,
	author = "Maggiore, Michele",
	title = "{Gravitational Waves. Vol. 1: Theory and Experiments}",
	doi = "10.1093/acprof:oso/9780198570745.001.0001",
	isbn = "978-0-19-171766-6, 978-0-19-852074-0",
	publisher = "Oxford University Press",
	year = "2007"
}

@article{Bluhm:2007bd,
	author = "Bluhm, Robert and Fung, Shu-Hong and Kostelecky, V. Alan",
	title = "{Spontaneous Lorentz and Diffeomorphism Violation, Massive Modes, and Gravity}",
	eprint = "0712.4119",
	archivePrefix = "arXiv",
	primaryClass = "hep-th",
	reportNumber = "IUHET-509",
	doi = "10.1103/PhysRevD.77.065020",
	journal = "Phys. Rev. D",
	volume = "77",
	pages = "065020",
	year = "2008"
}

@article{Schumacher:2023jxq,
	author = "Schumacher, Kristen and Yunes, Nicolas and Yagi, Kent",
	title = "{Gravitational wave polarizations with different propagation speeds}",
	eprint = "2308.05589",
	archivePrefix = "arXiv",
	primaryClass = "gr-qc",
	doi = "10.1103/PhysRevD.108.104038",
	journal = "Phys. Rev. D",
	volume = "108",
	number = "10",
	pages = "104038",
	year = "2023"
}

@article{Schumacher:2025plq,
	author = "Schumacher, Kristen and Talbot, Colm and Holz, Daniel E. and Yunes, Nicol{\'a}s",
	title = "{New test for superluminal gravitational wave polarizations}",
	eprint = "2501.18125",
	archivePrefix = "arXiv",
	primaryClass = "gr-qc",
	doi = "10.1103/m7ns-f3jf",
	journal = "Phys. Rev. D",
	volume = "112",
	number = "2",
	pages = "024067",
	year = "2025"
}

@article{Newman:1961qr,
	author = "Newman, Ezra and Penrose, Roger",
	title = "{An Approach to gravitational radiation by a method of spin coefficients}",
	doi = "10.1063/1.1724257",
	journal = "J. Math. Phys.",
	volume = "3",
	pages = "566--578",
	year = "1962"
}

@article{Hyun:2018pgn,
	author = "Hyun, Young-Hwan and Kim, Yoonbai and Lee, Seokcheon",
	title = "{Exact amplitudes of six polarization modes for gravitational waves}",
	eprint = "1810.09316",
	archivePrefix = "arXiv",
	primaryClass = "gr-qc",
	doi = "10.1103/PhysRevD.99.124002",
	journal = "Phys. Rev. D",
	volume = "99",
	number = "12",
	pages = "124002",
	year = "2019"
}

@article{Hohmann:2018wxu,
	author = "Hohmann, Manuel and Pfeifer, Christian and Levi Said, Jackson and Ualikhanova, Ulbossyn",
	title = "{Propagation of gravitational waves in symmetric teleparallel gravity theories}",
	eprint = "1808.02894",
	archivePrefix = "arXiv",
	primaryClass = "gr-qc",
	doi = "10.1103/PhysRevD.99.024009",
	journal = "Phys. Rev. D",
	volume = "99",
	number = "2",
	pages = "024009",
	year = "2019"
}

@article{Gong:2018ybk,
	author = "Gong, Yungui and Hou, Shaoqi",
	editor = "Calcagni, Gianluca",
	title = "{The Polarizations of Gravitational Waves}",
	eprint = "1806.04027",
	archivePrefix = "arXiv",
	primaryClass = "gr-qc",
	doi = "10.3390/universe4080085",
	journal = "Universe",
	volume = "4",
	number = "8",
	pages = "85",
	year = "2018"
}

@article{Gomes:2025ers,
	author = "Gomes, Cl{\'a}udio and Rosa, Jo{\~a}o Lu{\'\i}s and Pinto, Miguel A. S.",
	title = "{Gravitational wave propagation in generalized hybrid metric-Palatini gravity}",
	eprint = "2506.12870",
	archivePrefix = "arXiv",
	primaryClass = "gr-qc",
	doi = "10.1140/epjc/s10052-025-15085-x",
	journal = "Eur. Phys. J. C",
	volume = "85",
	number = "11",
	pages = "1359",
	year = "2025"
}

@article{DeFelice:2024dbj,
	author = "De Felice, Antonio and Mukohyama, Shinji and Tsujikawa, Shinji and Wang, Anzhong and Zhang, Chao",
	title = "{Even- and odd-parity stabilities of black holes in Einstein-Aether gravity}",
	eprint = "2407.00287",
	archivePrefix = "arXiv",
	primaryClass = "gr-qc",
	reportNumber = "YITP-24-74, IPMU24-0028, WUCG-24-06",
	doi = "10.1088/1475-7516/2024/08/056",
	journal = "JCAP",
	volume = "08",
	pages = "056",
	year = "2024"
}

@article{Zhang:2024cbw,
	author = "Zhang, Chao and Kase, Ryotaro",
	title = "{Even-parity stability of hairy black holes in U(1) gauge-invariant scalar-vector-tensor theories}",
	eprint = "2404.11910",
	archivePrefix = "arXiv",
	primaryClass = "gr-qc",
	doi = "10.1103/PhysRevD.110.044047",
	journal = "Phys. Rev. D",
	volume = "110",
	number = "4",
	pages = "044047",
	year = "2024"
}

@article{Ghilencea:2026jgp,
	author = "Ghilencea, D. M. and Mandric, V. -M.",
	title = "{Weyl gauge symmetry at LIGO-Virgo-KAGRA}",
	eprint = "2607.13146",
	archivePrefix = "arXiv",
	primaryClass = "gr-qc",
	month = "7",
	year = "2026"
}

@article{Dirac:1950pj,
	author = "Dirac, Paul A. M.",
	title = "{Generalized Hamiltonian dynamics}",
	doi = "10.4153/CJM-1950-012-1",
	journal = "Can. J. Math.",
	volume = "2",
	pages = "129--148",
	year = "1950"
}

\end{document}